\documentclass[10pt,journal,compsoc]{IEEEtran}
\usepackage{adjustbox}
\usepackage{caption}
\usepackage{booktabs}
\usepackage{wrapfig}
\usepackage{cite}
\usepackage{amsmath,amssymb,amsfonts}
\usepackage{algorithmic}
\usepackage{graphicx}
\usepackage{textcomp}
\usepackage{xcolor}
\usepackage{balance}
\usepackage{algorithm}
\usepackage{arevmath}
\usepackage{multicol}
\usepackage{multirow}
\usepackage{booktabs}
\usepackage{tabularx}
\usepackage{longtable}
\usepackage{array}
\usepackage{threeparttable}
\usepackage{colortbl}
\usepackage{enumitem}
\usepackage{hyperref} 
\usepackage{colortbl}
\usepackage{xspace}
\usepackage{subcaption}
\usepackage{float}
\usepackage{tikz}
\usepackage[most]{tcolorbox}

\newcommand{\sectopic}[1]{\vspace{0em}\par\noindent{\textit{\bfseries #1}}}

\begin{document}

\title{NLP-based Automated Compliance Checking of Data Processing Agreements against GDPR}

\author{Orlando~Amaral,~\IEEEmembership{Member,~IEEE,}
        Muhammad~Ilyas~Azeem,~\IEEEmembership{Member,~IEEE,}
        Sallam Abualhaija,~\IEEEmembership{Member,~IEEE,}
        and~Lionel~C~Briand,~\IEEEmembership{Fellow,~IEEE}
\IEEEcompsocitemizethanks{
\IEEEcompsocthanksitem This work has been submitted to the IEEE for possible publication. Copyright may be transferred without notice, after which this version may no longer be accessible.
\IEEEcompsocthanksitem O.Amaral, M.I. Azeem, S. Abualhaija, and L.C. Briand are with the SnT Centre for Security, Reliability, and Trust, University of Luxembourg, Luxembourg.\protect\\
E-mail: \{orlando.amaralcejas, ilyas.azeem, sallam.abualhaija, lionel.briand\}@uni.lu
\IEEEcompsocthanksitem L.C. Briand is also affiliated with the school of Electrical Engineering and Computer Science, University of Ottawa, Canada.\protect\\
E-mail: \{lbriand\}@uottawa.ca.
}
\thanks{Manuscript received Month DD, 2022; revised Month DD, 2022.}}

\markboth{Journal of \LaTeX\ Class Files,~Vol.~14, No.~8, Month~2022}%
{Amaral \MakeLowercase{\textit{et al.}}: NLP-based Automated Compliance Checking of Data Processing Agreements against GDPR}

\IEEEtitleabstractindextext{

\begin{abstract}
\textcolor{black}{When the entity processing personal data (the processor) differs from the one  collecting personal data (the controller), processing personal data is regulated in Europe by the General Data Protection Regulation (GDPR) through \textit{data processing agreements (DPAs)}.}
Checking the compliance of DPAs contributes to the compliance verification of software systems as DPAs are an important source of requirements for software development involving the processing of personal data. 
However, manually checking whether a given DPA complies with GDPR is challenging as it requires significant time and effort for understanding and identifying DPA-relevant compliance requirements in GDPR and then verifying these requirements in the DPA. Legal texts introduce additional complexity due to convoluted language and inherent ambiguity leading to potential misunderstandings.
In this paper, we propose an automated solution to check the compliance of a given DPA against GDPR. In close interaction with legal experts, we first built two artifacts: (i) the ``shall'' requirements extracted from the GDPR provisions relevant to DPA compliance and (ii) a glossary table defining the legal concepts in the requirements. Then, we developed an automated solution that leverages natural language processing (NLP) technologies to check the compliance of a given DPA against these ``shall'' requirements. Specifically, our approach automatically generates phrasal-level representations for the textual content of the DPA and compares them against predefined representations of the ``shall'' requirements. 
By comparing these two representations, the approach not only assesses whether the DPA is GDPR compliant but it further provides recommendations about missing information in the DPA. \textcolor{black}{Over a dataset of 30 actual DPAs, the approach correctly finds 618 out of 750 genuine violations while raising 76 false violations, and further correctly identifies 524 satisfied requirements. The approach has thus an average precision of 89.1\%, a recall of 82.4\%, and an accuracy of 84.6\%}. Compared to a baseline that relies on off-the-shelf NLP tools, our approach provides an average accuracy gain of $\approx$20 percentage points. The accuracy of our approach can be improved to $\approx$94\% with limited manual verification effort.
\end{abstract}

}

\maketitle
\IEEEdisplaynontitleabstractindextext

\IEEEpeerreviewmaketitle

\section{Introduction} \label{sec:introduction}

\textcolor{black}{
The advances in AI technologies led to increasingly integrated modern software systems (e.g., mobile applications) in everyday life activities. Applications are often developed to provide convenient services or assistance to individuals (e.g., Amazon Alexa). However, they also entail challenges regarding compliance to privacy standards and data protection requirements, e.g., when sharing personal data with third parties~\cite{zhan:22}. Individuals (i.e., the applications' end-users) who provide explicit consent to these applications for allowing the processing of their personal data must have guarantees that their data remains protected, also when shared with third parties~\cite{feal:20}. 
In Europe and even worldwide, the general data protection regulation (GDPR) was put into effect in 2018 to oblige  organizations to provide such data protection guarantees when handling personal data of European residents. Violating GDPR can levy penalty fines reaching up to tens of millions of euros~\cite{GDPR2018}. Software systems which typically involve processing or sharing personal data are subject to compliance with GDPR. 
Developing GDPR-compliant software requires taking into account the legally-binding agreements signed between different organizations involved in collecting and processing personal data. Examples of such agreements include privacy policies between organizations and individuals (often signed by individuals) as well as \textit{data processing agreements (DPAs)} between two organizations involved in collecting and processing personal data. Individuals are not necessarily aware of DPAs. We focus in our work on verifying the compliance of DPAs against the provisions of GDPR. 
From a requirements engineering (RE) standpoint, this compliance verification is a prerequisite for capturing a complete set of \textit{privacy-related requirements} covered in the DPAs to be implemented in software systems for processing personal data.}

Data processing normally involves an individual (i.e., \textit{data subject}) who willingly shares personal data, an organization (i.e., \textit{data controller}) that collects and in many cases further shares personal data, and another organization (i.e., \textit{data processor}) that processes personal data on behalf of the controller.
A data processor can share personal data again with yet another organization (i.e., \textit{sub-processor}) to perform some data processing services on its behalf. 
Individuals in this chain are often aware of the terms based on which their personal data is collected and handled as described in privacy notices of the data controller.  
All subsequent sharing of individual's personal data with processors (and sub-processors) is however not directly visible to individuals. According to GDPR, both the controller and processor should share the responsibility of protecting personal data of individuals. Consequently, a DPA listing privacy-related requirements should be established between the controller and the data processor(s)~\cite{pantlin:18}. 

DPAs are legally binding agreements that regulate the data processing activities according to GDPR. To be deemed GDPR-compliant, the DPA must explicitly cover all the criteria imposed by the GDPR provisions concerning data processing. Establishing a DPA includes setting terms for how data is used, stored, protected, and accessed as well as defining the obligations and rights of the controller and processor. By signing a DPA, the processor is obliged to ensure that any software system deployed for processing personal data has to also comply with GDPR. To illustrate, consider the scenario of an educational institution (called as \textit{Sefer University}) which shares personal data of its employees with a third-party service provider (e.g., accounting office) for a particular purpose (e.g., payroll administration). Examples of shared personal data include: first and last name, birth date, marital status, annual summary of leave or sickness absences and a scanned copy of a valid personal identification. The data controller in this scenario is Sefer University which collects the personal data of the employees (data subjects). The data processor is the service provider (in our example, the accounting office) performing financial services on behalf of the controller.  
Details concerning this service provider (e.g., the name and address of the hired accounting office) might not be shared with employees. To ensure that their personal data is sufficiently protected, GDPR requires Sefer university to sign a DPA with the accounting office. 

\begin{figure*}[!t]
\centering
\includegraphics[width=0.98\linewidth] {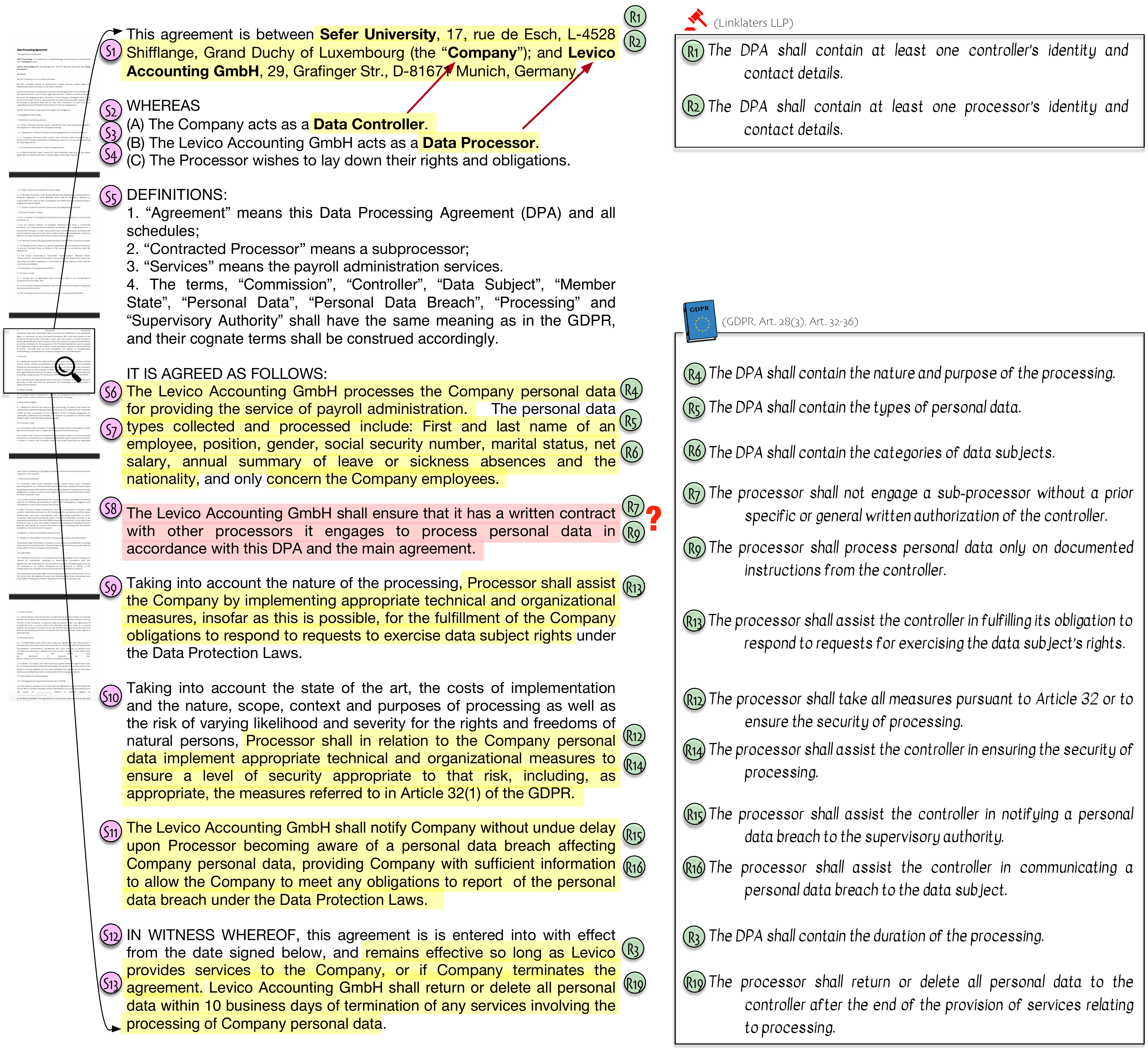}
\caption{Excerpt from a DPA verified against GDPR requirements.} \label{fig:exampledpa}
\vspace{-1em}
\end{figure*}

\subsection{Running Example } Fig.~\ref{fig:exampledpa} shows on the left-hand side an excerpt from a DPA that addresses the scenario above, and on the right-hand side a set of DPA-related requirements extracted from GDPR, ordered by their occurrence in the DPA. In Section~\ref{sec:qualitative}, we elaborate on the complete list of GDPR requirements. In the remainder of the paper, we use \textit{requirements} to refer to the privacy-related requirements extracted from GDPR concerning data processing, and \textit{statements} to refer to the textual content of a given DPA. For simplicity, a statement is regarded as equivalent to one sentence in the DPA. 

Checking the compliance of the DPA in Fig.~\ref{fig:exampledpa} requires mapping the statements (marked as S1 -- S13) in the DPA against the requirements envisaged by GDPR (marked as R1 -- R7, R9, R12 -- R16, and R19). Note that the requirements numbering is not sequential and is defined by the complete list presented in Table~\ref{tab:mandatory-requirements}. The most immediate straight-forward solution that comes to mind for matching statements with requirements is by applying semantic similarity measures, e.g., cosine similarity~\cite{Jurafsky:09}. In brief, cosine similarity is a measure that assigns a score between 0 and 1 to a pair of text sequences. The higher the score, the more semantically similar the two text sequences. Following this, a statement is satisfying a given requirement when the cosine similarity is sufficiently high.  
Below, we discuss the limitations of such a simple solution. 

The example shows that S1 satisfies both R1 and R2 since it clearly identifies the data controller (Sefer University) and processor (Levico Accounting GmbH). However, computing the semantic similarity between S1 and R1 and R2 results in scores equal to 0.02 and 0.14, respectively. Low similarity scores in this case should be expected since S1 contains specific entities in place of the identity and contact details of the controller and processor (required by GDPR). Without considering S2 and S3, one cannot  distinguish between the data controller and processor fully automatically. Since this is key information that has a significant impact on compliance checking, we assume in our work that the name(s) of the data controller and processor are provided as input by the human analyst. To properly verify the compliance of a given DPA, the requirements have to be checked for the right entity. For example, the processor's obligations in R7, R9, R12-R16, and R19 in Fig.~\ref{fig:exampledpa} must be  associated with the processor. On a similar basis, S12 has a similarity score of 0.3 with R3, S6 0.38 with R4, S7 0.53 with R5 and 0.33 with R6. All these semantic similarity scores are too low to enable a decision about compliance. Using semantic similarity is not suitable for checking the compliance against R1 -- R6 since they require content that varies across DPAs, e.g., a list of collected personal data or the processing purpose. 

The example further shows that S8 satisfies neither R7 nor R9, yet it has a semantic similarity of 0.52 and 0.57, respectively. S8 is about having a written agreement with sub-processors, while R7 concerns acquiring a written approval from the controller before engaging sub-processors. R9 is related to processing personal data only based on documented instructions from the controller. Though there is some resemblance in the wording of S8 and that of R7 and R9, it is still clear that S8 does not comply with either requirements. 

The example illustrates that statements in DPAs are often verbose, i.e., include more content that is needed for GDPR compliance.  For example, S9 satisfies R13 with a similarity score of 0.87, whereas S10 satisfies both R12 and R14 with a similarity score of 0.52. This indicates that irrelevant content in S10 reduces the overall semantic similarity, and thus impedes compliance verification. Computing the similarity with only the relevant part of S10 (shaded yellow) leads to higher scores: 0.71 with R12 and 0.60 with R14. 

Further, statements can satisfy multiple requirements. For example, S11 satisfies R15 and R16. The similarity scores between S11 and R15 and R16 are 0.82 and 0.79, respectively. However, S11 does not specify stating to whom data breaches shall be notified (that is the supervisory authority in R15 or the data subject in R16). In this case, S11 can be regarded as \textit{partially} satisfying both requirements. We see that a GDPR requirement can contain different key elements which must be verified in the DPA statement. For instance, key elements in R15 include \textit{the processor}, \textit{assist in notifying}, \textit{the controller}, \textit{personal data breach}, and \textit{supervisory authority}. Verifying all key elements is necessary to understand the missing content in a statement. To become compliant, S11 can be improved by adding missing information content.

The above discussion highlights that the seemingly simple task of mapping statements in a given DPA to GDPR requirements cannot be addressed accurately through straight-forward automation based on semantic similarity. Alternatively, manually analyzing the content of DPAs for compliance against GDPR is a very tedious and time-consuming task for legal experts and requirements engineers alike. In this paper, we propose a natural language processing (NLP)-based automation for checking the compliance of DPAs against GDPR. We develop our solution in close collaboration with legal experts from our industry partner (Linklaters LLP - a multinational law firm). 
\textcolor{black}{Linklaters has top-tier rankings across many legal-practice areas, including funds investment, banking, and
finance. The legal experts in Linklaters have long experience in providing GDPR advisory and compliance services.}

Ensuring that the DPA excerpt in Fig.~\ref{fig:exampledpa} is compliant with GDPR, and consequently provides a  complete list of requirements, has an impact on building a GDPR-compliant software system. For example, a system deployed by the processor shall not process or store any personal data outside the list specified in S7. The system shall further implement: (i) appropriate protection procedures (e.g., encryption, data anonymization) to ensure the required level of security in accordance with S10, and (ii) a data breach notification procedure, corresponding to S11, which shall send a text message to the controller and also collect an acknowledgment of receipt.

As we elaborate in Section~\ref{sec:related}, much work has been done in the RE literature on automating regulatory compliance. Despite being an essential source for compliance requirements of data processing activities in software systems, automated compliance checking of DPAs against GDPR has not been previously studied in RE. 
Moreover, none of the existing work leverages the key elements of requirements in GDPR (e.g., the case of R15 and R16 discussed above) for a phrasal-level, automated compliance checking. Analyzing the content of DPAs at a phrasal level is beneficial for understanding the exact missing information content in the DPA and hence recommending remedies to prevent non-compliance with GDPR.  

\begin{figure} 
\centering
\includegraphics[width=0.48\textwidth]{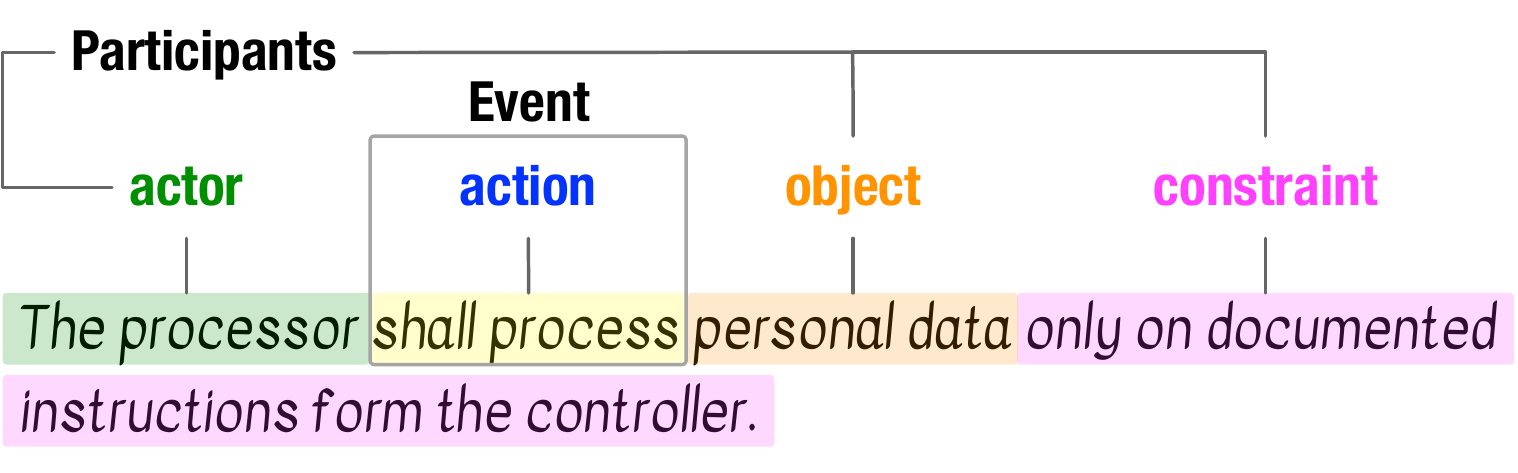}
\vspace{-.5em}
\caption{Illustration of SF-based representation.}
\label{fig:sf-eg}
\vspace{-1.2em}
\end{figure}

In our work, we propose using semantic frames~\cite{Jurafsky:09} for generating an intermediate representation to characterize the information content in each GDPR requirement and thus enable a meaningful, phrasal-level decomposition of that requirement. As we elaborate in Section~\ref{sec:background}, a \textit{semantic frame (SF)} of a given \textcolor{black}{sentence} describes an event and a set of participants, where each participant has a specific \textit{semantic role (SR)} in that event. Fig.~\ref{fig:sf-eg} illustrates SF elements describing  a \textit{processing} event in a requirement example extracted from GDPR Art.~28(3)(a)  -- corresponding to R9 in Fig.~\ref{fig:exampledpa}. The event \textit{processing}, identified through the SR \textit{action}, evokes three participants: ``the processor'' who performs processing (i.e., \textit{actor}), ''personal data'' on which processing is performed (i.e.,  \textit{object}) and ``only on documented instructions from the controller'' that restricts processing (i.e., \textit{constraint}). 

\subsection{Research Questions } 
This paper investigates four Research Questions (RQs):

\sectopic{RQ1: What are the requirements for checking the compliance of a DPA according to GDPR?} 
We answer RQ1 by defining a set of 45 ``shall'' requirements concerning the compliance of a DPA. 
Verifying these requirements in DPAs  ensures that the DPAs contain a complete set of necessary legal requirements, to be implemented in software systems, concerning data processing activities. We further create a glossary table that defines the legal concepts related to DPA compliance and provide traceable links to GDPR articles. We present the compliance requirements in Section~\ref{sec:qualitative}.

\sectopic{RQ2: How can a DPA be automatically checked for compliance against GDPR at a phrasal level?} 
To address RQ2, we first manually define SF-based representations for the 45 compliance requirements from RQ1. Then, we devise an automated approach that leverages various NLP technologies to automatically generate SF-based representations for the textual content of a given DPA. Our approach then verifies the compliance of the DPA by aligning its respective SF-based representations against the representation of each GDPR requirement. 
We elaborate our approach in Section~\ref{sec:approach}. 

\sectopic{RQ3: How accurately can we check the compliance of a given DPA?} 
RQ3 assesses the accuracy of our approach in checking the compliance of DPAs according to GDPR provisions.  
\textcolor{black}{Over an evaluation set of 30 real DPAs, our approach successfully detects 618 out of 750 violations and introduces false violations in 76 cases. Our approach has thus a precision of 89.1\% and a recall of 82.4\%}.
\textcolor{black}{Compared to a baseline that employs off-the-shelf NLP tools, our approach yields on average a gain of $\approx$20 percentage points in accuracy, and $\approx$17 percentage points in both precision and recall.} We discuss RQ3 and the subsequent RQs in Section~\ref{sec:evaluation}.

\sectopic{RQ4: How efficient is our approach in terms of execution time?}
Based on our experience, reviewing entirely manually the content of a given DPA containing about 200 statements takes on average 30 minutes if one is intimately familiar with the GDPR articles relevant to data processing activities. In contrast, our approach automatically analyzes a DPA of the same size in $\approx$2.5 minutes, and further produces as output a detailed report which summarizes the results, including the compliance decision together with the missing information and the respective recommendations to achieve compliance.

\subsection{Contributions } 
We take steps toward addressing the limitations outlined above. Concretely, our contributions are as follows:

(1) We extract in close interaction with subject-matter experts, a list of 45 compliance requirements from the GDPR provisions concerning data processing. We further simplify and document these compliance requirements as ``shall'' requirements. Our motivation is to increase readability for requirements engineers by using a familiar format, while maintaining usability for legal experts since these requirements rely on GDPR terminology.

(2) We devise an automated approach that leverages NLP technologies for automatically checking the compliance of DPAs against GDPR. More specifically, we rely on syntactic and semantic parsing to generate intermediate SF-based representations that are then compared for compliance checking. Our proposed SF-based representation is based on a set of semantic roles with intuitive descriptions. 
Future changes in compliance requirements is a common concern for legal experts and requirements engineers. With enough training, additional compliance requirements can be added to our automated approach by defining their respective SF-based representations. The SF-based representation is used as an enabler for automated compliance checking. 

(3) We empirically evaluate our approach on 54 real DPAs. \textcolor{black}{We randomly split the DPAs into two mutually exclusive subsets as follows. 
The fist subset containing 24 DPAs was used for creating the SF-based representations and developing our approach, whereas  the second subset, including the remaining 30 DPAs, was used exclusively for evaluation.} 
The evaluation set collectively contains a total of 7048 statements manually reviewed and analyzed for compliance by third-party annotators. \textcolor{black}{On this evaluation set, our approach yields a precision of 89.1\%, and a recall of 82.4\%}. Our approach outperforms a straight-forward baseline that employs NLP off-the-shelf tools with a percentage point gain of $\approx$17 in both precision and recall. 

{\color{black}(4) We propose computing and assigning scores to requirements found to be satisfied in a given DPA. Such score indicates the maximum matching degree between a requirement and any DPA statement satisfying it. The matching degree represents the ratio of key elements that are present in a DPA statement compared to the elements to be expected given the requirement. We discuss how matching degrees are computed in detail in Section~\ref{sec:approach}. These scores are intended to facilitate the review and validation of findings from our automated approach by a human analyst. For example, our empirical evaluation shows that the analyst can improve the overall accuracy of our automated approach by 6.4 percentage points when reviewing statements satisfying requirements with the maximum matching degree scores as long as the scores are less than or equal to a threshold of 0.5. The proportion of these statements, on average, is $\approx$6\% (corresponding to an average of \textcolor{black}{12 statements} in a DPA). This accuracy improvement comes with a gain of 1.2 and 11.5 percentage points in precision and recall, respectively.  
}

\subsection{Structure.} Section~\ref{sec:background} explains the necessary background for our approach. Section~\ref{sec:qualitative} presents how we extracted the DPA-related requirements from GDPR. Section~\ref{sec:approach} describes our automated approach. Section~\ref{sec:evaluation} evaluates our approach. Section~\ref{sec:related} reviews related work. Section~\ref{sec:threats} discusses threats to validity. \textcolor{black}{Section~\ref{sec:reusability} describes how our methodology can be adapted beyond GDPR or DPAs.} Finally, Section~\ref{sec:conclusion} concludes the paper.

\section{Background} \label{sec:background}

In this section, we briefly introduce GDPR, and explain the NLP background on semantic frames.    
\subsection{GDPR} 
GDPR~\cite{EU2018} is the benchmark for data privacy regulations in Europe with 173 recitals and 99 articles divided into 11 chapters. GDPR applies primarily to organizations in Europe, but it can also impose obligations on organizations outside Europe when they offer goods or services, or monitor individuals in Europe. According to GDPR, there are often two main actors that should be identified in agreements related to processing personal data, namely a data controller and data processor.  A data controller determines the purpose and means of processing personal data, whereas a data processor processes personal data based on the instructions of the controller. GDPR imposes various obligations on a given organization when it is acting as a processor or controller, the latter being subject to more provisions than the former. 

Data processors are obliged to: (1)~implement adequate technical and organizational measures to ensure the security of personal data, and, in case of data breaches, notify the controllers without undue delay; (2)~appoint a statutory data protection officer (if needed) and conduct a formal data protection impact assessment for certain types of high-risk processing; (3)~keep records about their data processing; and (4)~comply to GDPR restrictions when transferring personal data outside Europe. Data controllers have to meet the obligations listed above, and further: (1)~adhere to six core personal data processing principles, namely, fair and lawful processing, purpose limitation, data minimization, data accuracy, storage limitation, and data security; (2)~keep identifiable individuals informed about how their personal data will be used; and (3)~ensure the individual rights envisaged by GDPR, e.g., the right to be forgotten and the right to lodge a complaint. 

GDPR includes some specific provisions in relation to DPAs, which in turn can have an impact on software development. For example, GDPR requires data processors to return or delete personal data without undue delay after the processing service is completed. For some cloud service providers, implementing this requirement, where immediate deletion is not possible, requires archiving personal data first until the next delete command is performed. We elaborate the GDPR requirements concerning DPAs in Section~\ref{sec:qualitative}.

\subsection{Semantic Frames} 
\label{subsec:semanticframes}
A semantic frame (SF) describes a set of elements that participate in a particular event~\cite{Jurafsky:09}. Each element in the frame is a text span that is labeled with a semantic role (SR), denoted as \textbf{[span]$_\textit{role}$}. 
For instance, the SF about a \textit{purchase event} in the \textcolor{black}{sentence} ``[William]$_\textit{buyer}$ purchased [a brand new car]$_\textit{purchased item}$'' contains the SRs  \textit{buyer} and  \textit{purchased item} spanning ``William'' and ``a brand new car'', respectively. SFs are often represented as \textit{predicate-arguments structure}, where a predicate describes an event represented by the text span of the verb, and each argument describes a participant in the event labeled with a specific SR. A predicate evokes a set of expected arguments based on the event. For example, the above \textcolor{black}{sentence} contains the arguments \textit{buyer} and \textit{purchased item}, but it could also contain \textit{seller} and \textit{price}.
The NLP community has long been creating machine-readable lexical resources like FrameNet~\cite{Baker:98} and PropBank~\cite{Kingsbury:02} that contain manually-labeled \hbox{predicates and arguments. }

One way to identify the SF of a given \textcolor{black}{sentence} is through semantic role labeling (SRL), which is the task of identifying the different SRs occurring in the \textcolor{black}{sentence}~\cite{Gildea:02}. 
In SRL, one identifies the text spans that provide answers to questions about who (i.e., \textit{actor}) did what (i.e., \textit{action}) to whom (\textit{beneficiary}), when (i.e., \textit{time}) and why (i.e., \textit{reason})~\cite{Li:21, Marquez:08}. 
SRL can be employed to improve the solutions of many downstream NLP tasks such as machine translation~\cite{Shi:16}, question answering~\cite{Berant:13, Shen:07} and relation extraction~\cite{Lin:17,Shi:19}. 
For instance, question answering can be improved by aligning the SRs in the question with those in the likely answer. If we identify the SRs in the question ``[What]$_\textit{purchased item}$ did [Mazda]$_\textit{seller}$ [sell]$_\textit{action}$ to [William]$_\textit{buyer}$?'', extracting the answer from the above example (i.e., ``a brand new car'') becomes straightforward.
Inspired by the NLP literature~\cite{Shen:07,Wu:11,Woodsend:14}, we check the compliance of a given DPA against GDPR requirements by generating SF-based representations.  In particular, we use the SR \textit{action} to define the predicate of an SF, and a set of other SRs (e.g., \textit{actor} and \textit{reason}) to define the arguments as we elaborate in Section~\ref{sec:approach}.

\section{Compliance Requirements for Data Processing in GDPR (\textbf{RQ1})}
\label{sec:qualitative}

In this section, we present two artifacts to answer \textbf{RQ1}, namely (1) a total of 45 compliance requirements for DPA according to GDPR; and (2) a glossary table defining the legal concepts in these requirements, including traceability to GDPR articles. The artifacts are provided in Table~\ref{tab:mandatory-requirements} as well as in the tables provided as supplementary material.

\begin{table*}[t]
\caption{\textcolor{black}{Mandatory DPA Compliance Requirements in GDPR.}}\label{tab:mandatory-requirements}
\vspace{-.5em}
 \footnotesize
  \centering
	  \begin{threeparttable}[t]
  \begin{tabularx}{0.98\textwidth} {@{} p{0.08\textwidth}  
  @{\hskip 0.5em}  X}
   \toprule
   ID (\textbf{Cat}\tnote{1}) & Requirement (\textit{Reference}\tnote{2}) \\ 
    \midrule
    R1 (\textbf{MD1}) & The DPA shall contain at least one controller's identity and contact details. (\textit{Linklaters LLP}) \\
        \midrule
    R2 (\textbf{MD2})& The DPA shall contain at least one processor's identity and contact details. (\textit{Linklaters LLP}) \\
    \midrule
    R3 (\textbf{MD3})& The DPA shall contain the duration of the processing. (\textit{Art.~28(3)})\\
    \midrule
    R4 (\textbf{MD4})&  The DPA shall contain the nature and purpose of the processing. (\textit{Art.~28(3)}) \\  
    \midrule
   R5 (\textbf{MD5})& The DPA shall contain the types of personal data. (\textit{Art.~28(3)})\\ 
    \midrule
    R6 (\textbf{MD6})& The DPA shall contain the categories of data subjects. (\textit{Art.~28(3)})\\
   \midrule
    R7 (\textbf{PO1})& The processor shall not engage a sub-processor without a prior specific or general written authorization of the controller. (\textit{Art.~28(2)}) \\ 
    \midrule
    R8 (\textbf{PO2})& In case of general written authorization, the processor shall inform the controller of any intended changes concerning the addition or replacement of sub-processors. (\textit{Art.~28(2)}) \\
    \midrule
    R9 (\textbf{PO3})& The processor shall process personal data only on documented instructions from the controller. (\textit{Art.~28(3)(a)})\\
    \midrule
    R10 (\textbf{PO4})& If the processor requires by Union or Member State law to process personal data without instructions and law does not prohibit informing the controller on grounds of public interest, the processor shall inform the controller of that legal requirement before processing. (\textit{Art.~28(3)(a)}) \\
    \midrule
    R11 (\textbf{PO5})& The processor shall ensure that persons authorized to process personal data have committed themselves to confidentiality or are under an appropriate statutory obligation of confidentiality. (\textit{Art.~28(3)(b)})\\
    \midrule
    R12 (\textbf{PO6})& The processor shall take all measures required pursuant to Article 32 or to ensure the security of processing. (\textit{Art.~28(3)(c)}) \\ 
    \midrule
    R13 (\textbf{PO7})& The processor shall assist the controller in fulfilling its obligation to respond to requests for exercising the data subject's rights. (\textit{Art.~28(3)(e)}) \\ 
      \midrule
      R14 (\textbf{PO8})& The processor shall assist the controller in ensuring the security of processing. (\textit{Art.~28(3)(f), Art.~32}) \\ 
      \midrule
      R15 (\textbf{PO9})&  The processor shall assist the controller in notifying a personal data breach to the supervisory authority. (\textit{Art.~28(3)(f), Art.~33})\\ 
      \midrule
      R16 (\textbf{PO10})&  The processor shall assist the controller in communicating a personal data breach to the data subject. (\textit{Art.~28(3)(f), Art.~34}) \\ 
      \midrule
      R17 (\textbf{PO11})&  The processor shall assist the controller in ensuring compliance with the obligations pursuant to data protection impact assessment (DPIA). (\textit{Art.~28(3)(f), Art.~35}) \\ 
      \midrule
      \textcolor{black}{R18} (\textbf{PO12})& The processor shall assist the controller in consulting the supervisory authorities prior to processing where the processing would result in a high risk in the absence of measures taken by the controller to mitigate the risk. (\textit{Art.~28(3)(f), Art.~36})\\
      \midrule
      R19 (\textbf{PO13})& The processor shall return or delete all personal data to the controller after the end of the provision of services relating to processing. (\textit{Art.~28(3)(g)})\\ 
      \midrule
      R20 (\textbf{PO14})& The processor shall immediately inform the controller if an instruction infringes the GDPR or other data protection provisions. (\textit{Art.~28(3)(h)})  \\
  \midrule
  R21 (\textbf{PO15})& The processor shall make available to the controller information necessary to demonstrate compliance with the obligations Article 28 in GDPR. (\textit{Art.~28(3)(h)}) \\
  \midrule
  R22 (\textbf{PO16})& The processor shall allow for and contribute to audits, including inspections, conducted by the controller or another auditor mandated by the controller. (\textit{Art.~28(3)(h)}) \\ 
  \midrule
  R23 (\textbf{PO17})& The processor shall impose the same obligations on the engaged sub-processors by way of contract or other legal act under Union or Member State law. (\textit{Art.~28(4)})\\
  \midrule
  R24 (\textbf{PO18})& The processor shall remain fully liable to the controller for the performance of sub-processor's obligations. (\textit{Art.~28(4)}) \\
  \midrule
  R25 (\textbf{PO19})& When assessing the level of security, the processor shall take into account the risk of accidental or unlawful destruction, loss, alternation, unauthorized disclosure of or access to the personal data transmitted, stored or processed. (\textit{Art.~32(2)}) \\
  \midrule
  R26 (\textbf{CR1})& In case of general written authorization, the controller shall have the right to object to changes concerning the addition or replacement of sub-processors, after having been informed of such intended changes by the processor.  (\textit{Art.~8(2)}) \\ 
   \bottomrule
 \end{tabularx}
 	\begin{tablenotes}
 	\item[1] Cat is the category of the requirement, namely metadata (MD), processor's obligation (PO), controller's right (CR), and controller's obligation (CO).  
    \item[2] GDPR-related articles.
    \end{tablenotes}
    \end{threeparttable}
\end{table*}

In collaboration with legal experts from Linklaters, we extracted from GDPR a total of 45 compliance requirements that regulate DPAs. 
\textcolor{black}{A GDPR-compliant DPA provides assurance about the completeness of the requirements concerning data processing activities, which are essential for developing compliant software systems.}  According to the experts' feedback, there are 26 mandatory and 19 optional requirements. \textcolor{black}{The different requirements' types were discussed and agreed upon during several 
sessions.} 

Mandatory requirements (R1 -- R26, listed in Table~\ref{tab:mandatory-requirements}) are explicitly stipulated in GDPR, except R1 and R2 which were strongly recommended by legal experts. R1 and R2 provide key information concerning the identity and contact details of the data controller and processor based on which the compliance checking is performed. Optional requirements (R27 -- R45, see supplementary material) are derived from good practice in writing DPAs according to legal experts. Our automated support results in a non-compliance decision when a mandatory requirement is violated, and raises a warning when an optional requirement is violated.

We further group the compliance requirements into four categories, namely nine \textit{metadata} requirements (denoted as MD), 25 requirements concerning the \textit{processor's obligation} (denoted as PO), three requirements about the \textit{controller's rights} (denoted as CR), and eight about the \textit{controller's obligations} (denoted as CO). As shown in Table~\ref{tab:mandatory-requirements}, the majority of mandatory requirements describe the processor's obligations (corresponding to 19/25). 

To build our artifacts, we followed an iterative method that includes three steps. The first step was reading the articles of GDPR related to DPA. In the second step, we created the artifacts. \textcolor{black}{Finally, we validated these artifacts with legal experts.} Through a close interaction with these experts, we incrementally improved and refined the artifacts to get them in their final form. 
To mitigate fatigue, we conducted each validation session for a duration of approximately two hours adding up to an overall of 46 hours for all sessions. \textcolor{black}{Three legal experts participated in these joint sessions. Their  expertise spans several areas including corporate laws, commercial litigation, data protection laws and financial regulations, and information technology within the legal domain.} 
\textcolor{black}{During these sessions, we refined the requirements according to the legal experts' interpretation, thereby alleviating any ambiguity in the GDPR.} 
During these steps, we created the glossary terms (the second artifact provided as supplementary material) with traceability links to GDPR articles to facilitate the understanding of the requirements. 

\begin{figure*} 
\centering
\includegraphics[width=0.95\textwidth]{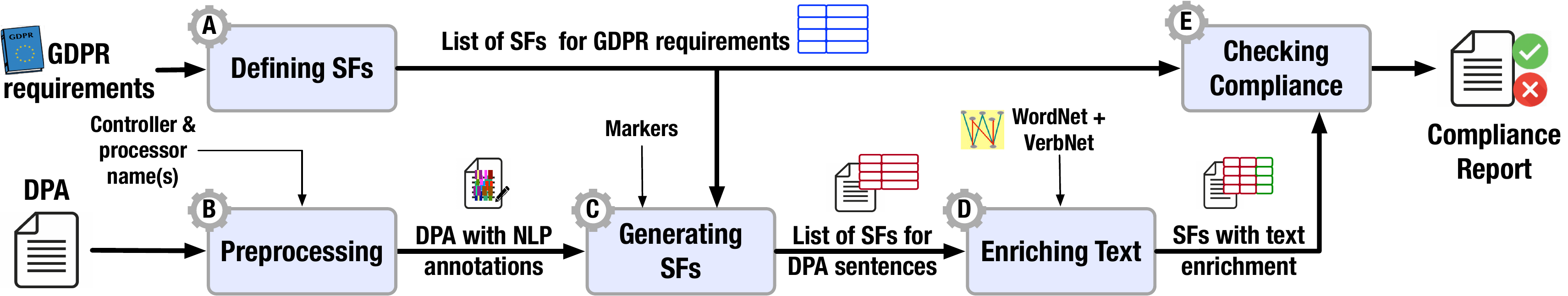}
\caption{Overview of our DPA compliance checking approach (\textit{DERECHA}).}
\label{fig:approach}
\end{figure*}

Following experts' recommendations, we started by analyzing Art.~28 in GDPR as it is directly relevant to the compliance of DPAs. We then read through all referenced articles in Art.~28, namely Articles 32–36, 40, 42, 43, 63, 82–84, and 93(2). 
While reading the articles, we (1) transformed the text in GDPR articles into ``shall'' requirements and (2) divided compound statements into simple requirements. 
For example, we extracted from Art.~28(3), which states that \textit{``Processing by a processor shall be governed by a contract [...] that sets out the subject-matter and duration of the processing, the nature and purpose of the processing, the type of personal data and categories of data subjects [...]''}, four requirements related to the metadata content of a DPA, marked as R3 -- R6 in Table~\ref{tab:mandatory-requirements}.
We adopted the GDPR text as-is when it resembled a ``shall'' requirement such as R34 (see supplementary material) 
extracted from Art.~33(2). 

Using the resulting list of 45 compliance requirements, we manually checked the compliance of 24 DPAs. Specifically, we identified for each requirements whether it is satisfied by any statement in the DPA. \textcolor{black}{As we elaborate in Section~\ref{subsec:docCollection}, our manual analysis aimed at establishing in-depth understanding of the compliance requirements before outsourcing the annotation to external annotators.}

\section{Automated Approach for DPA Compliance Checking against GDPR (\textbf{RQ2})} \label{sec:approach}

To address \textbf{RQ2}, we devise an NLP-based approach for checking the compliance of DPAs against GDPR at a phrasal level (thereafter referred to as \textit{DERECHA}, standing for \textbf{D}PA s\textbf{E}mantic f\textbf{R}am\textbf{E}-based \textbf{C}ompliance c\textbf{H}ecking \textbf{A}gainst GDPR). 
The current version of \textit{DERECHA} does not use machine learning or deep learning architectures as, in our application context, we typically do not have enough labeled data to train robust models with accurate predictions. 
Recent advances in the NLP literature show that fine-tuning large-scale language models like BERT~\cite{Devlin:18} often yields accurate models for many downstream tasks~\cite{jeremy:18, Jiang:19, Yates:21}. 
To this end, we note that such language models are integrated into currently available off-the-shelf semantic role labeling tools 
which DERECHA does not directly use as we explain later in this section. Instead, we employ these off-the-shelf NLP tools as a baseline solution 
and compare its performance in Section~\ref{subsec:evalProc} against the performance of \textit{DERECHA}. 
That said, we believe that attempting to extend our dataset and experimenting with machine learning or deep learning models would be important to refine our conclusions. This is left for future work.

Fig.~\ref{fig:approach} presents an overview of \textit{DERECHA}, composed of five steps. The input to \textit{DERECHA} includes a DPA, the name(s) of the controller and processor, and a set of requirements. The output is then a compliance report which summarizes the findings of \textit{DERECHA}.
In step~A, we manually create SF-based representations of the compliance requirements extracted in Section~\ref{sec:qualitative}. Note that this step is performed only once and can be reused across DPAs. In practice, from a user perspective, the SF-based representations identified in this paper can be adopted as-is.
In step~B, we apply an NLP pipeline to parse the DPA and preprocess the text. In step~C, we automatically generate SF-based representations for the input DPA. In step~D, we enrich the textual content in the DPA with semantically-related text, e.g., the text ``engage'' will be enriched with ``hire'' and ``employ''. In the last step, step~E, we use the SF-based representations created in steps~A~and~D to check the compliance of the input DPA against GDPR. Finally, we generate a detailed report about the compliance decision. We elaborate these steps next.

\subsection{Step~A: Defining SFs in Compliance Requirements} \label{subsec:stepa}
In this step, we define an SF-based representation to characterize the information content of each requirement in GDPR. This representation aims at decomposing the requirement and thus enabling compliance checking at a phrasal level.  

As can be seen in Table~\ref{tab:mandatory-requirements}, the metadata requirements describe the content that a DPA has to cover, e.g., the duration of the agreement. In contrast, the requirements under the other categories (e.g., processor's obligations) are expected to be precisely stated in the DPA.  While the exact terminology is not required by GDPR, the semantics of these requirements are. This has an implication on the compliance checking process. For example, a statement in a given DPA satisfying R5 (\textbf{MD5}) will unlikely mention ``the DPA'' or ``shall contain''. On the contrary, a statement satisfying R9 (\textbf{PO3}) has to explicitly cover the different elements of its SF-based representation illustrated in Fig.~\ref{fig:sf-eg}. Consequently, we define such representations only for the requirements concerning processor's obligations, controller's rights and controller's obligations. We elaborate on how we handle the compliance checking against the metadata requirements (R1 -- R6 and R27 -- R29) in Section~\ref{subsec:stepe}.

To define the SF-based representation, we first identify the SRs in each requirement. 
Inspired by existing literature~\cite{Gildea:02,Sleimi:18, Bhatia:19}, we collate a list of 10 SRs pertinent to DPA compliance requirements in GDPR. Four SRs are illustrated in Fig.~\ref{fig:sf-eg}, including \textit{action} to describe a particular event, and \textit{actor}, \textit{object}, and \textit{constraint} to describe the SRs of the participants in that event. 
The remaining six SRs, exemplified on R18 -- R20 in Table~\ref{tab:mandatory-requirements}, are: \textit{Beneficiary} -- an entity that benefits from an action (e.g., [the controller]$_\textit{beneficiary}$ in R18);  \textit{reason} -- justifying why the action is performed (e.g., [to mitigate the risk]$_\textit{reason}$ in R18);  \textit{time} -- a temporal aspect (e.g., [prior to processing]$_\textit{time}$ in R18); \textit{condition} -- the cases where the action is performed (e.g., [If an instruction infringes GDPR or other data protection provisions]$_\textit{condition}$ in R20); 
\textit{situation} -- something that happened or can happen (e.g., [the end of provision of services related to processing]$_\textit{situation}$ in R19); and finally \textit{reference} -- the mention of legal entities (e.g., [GDPR or other data protection provision]$_\textit{reference}$ in R20). 

Recall from Section~\ref{sec:background} that an SF is represented as a predicate-arguments structure. 
For each requirement, we therefore define a predicate and a set of arguments evoked by that predicate. 
The predicate highlights the main action a data processor needs to perform to be compliant with GDPR. Example predicates include $\langle \text{not engage}\rangle$ in R7, $\langle \text{process}\rangle$ in R9, and $\langle \text{return or delete}\rangle$ in R19. The arguments are the different phrases in the requirement which constitute the participants associated with the predicate.  We manually identify the arguments by applying the list of SRs explained above on the requirement text.
Different requirements can share the same predicate when they refer to the same action, e.g., six requirements share the predicate $\langle \text{assist}\rangle$, namely R13 -- R18. Nonetheless, we define one representation for each requirement, since the required arguments might be different. 
Doing so is essential for capturing missing information content in DPA statements according to what is expected by each requirement. 
Finally, we have discussed the SF-based representations with our collaborating experts from Linklaters and received feedback that the underlying SRs are intuitive and easy to understand. This suggests that extending our approach with more compliance requirements alongside their respective SF-based representations is practically feasible. 

The first step results in: (i) a list of SRs collated and refined from the existing literature which is passed on to step~C to be further used in automatically generating SF-based representations for the input DPA; and (ii) SF-based representations for the requirements (R7 -- R25 and R30 -- R45) passed on to step~E. These representations enable automated compliance verification of the statements at a phrasal level. An example of such representation for \textbf{PO6} (R12 in Table~\ref{tab:mandatory-requirements}) contains the predicate $\langle \text{\textbf{take}}\rangle$, and a set of four arguments $\mathcal{A}=\{a_i : 0\leq i \leq 3\}$ such that $a_0$~=~[the processor]$_\textit{actor}$, $a_1$~=~[all measures]$_\textit{object}$, $a_2$~=~[Article 32]$_\textit{reference}$, $a_3$~=~[to ensure the security of processing]$_\textit{reason}$. The SF-based representations for all 45 requirements are provided as supplementary material. 

\subsection{Step~B: Preprocessing} \label{subsec:stepb}
Given a DPA as input, this step applies an NLP pipeline 
composed of seven modules, including: 
(1) \textit{tokenization}, splits the text into tokens; (2) \textit{\textcolor{black}{sentence} splitting}, identifies the different sentences in the text using different heuristics, e.g., a \textcolor{black}{sentence} ends with a period; (3) \textit{lemmatization}, maps the words to their canonical forms, e.g., ``instructions'' and ``requested'' are mapped to ``instruction'' and ``request''; (4) \textit{POS tagging} assigns a part-of-speech (POS) tag to each token, e.g., noun, verb, and  possessive pronoun; (5) \textit{chunking}, groups the words that form a syntactic unit, e.g., ``the processor'' is a noun phrase; (6) \textit{dependency parsing}, describes the grammatical relations in a \textcolor{black}{sentence}, e.g., ``the processor'' and ``personal data'' in the \textcolor{black}{sentence} ``the processor transfers personal data'' are the subject and the object of the verb ``transfers'', respectively; and 
finally (7) \textit{semantic parsing}, defines the meanings of the different constituents in a \textcolor{black}{sentence}. For semantic parsing, we apply two widely-known lexical resources, namely WordNet~\cite{Miller:95, Miller:98} and VerbNet~\cite{Korhonen:04,Kipper:06}. 
WordNet is a lexical database for English. It groups words into sets of synonyms called \textit{synsets} and further connects the synsets via semantic relations. For example, \textit{is-a} connects a hyponym (more specific synset) to a hypernym (more general synset)~\cite{Agirre:07}. 
VerbNet groups verbs into classes according to their structure, e.g., the verbs under the \textit{motion} class include run, escape, and leave~\cite{Schuler:05}. 

Each module in this pipeline generates NLP annotations on parts of the text in the input DPA. The resulting NLP annotations are passed on to the next steps.

\subsection{Step~C: Generating SFs Automatically} \label{subsec:stepc}
In step~C, we automatically generate for each statement in the input DPA an SF-based representation. 
To do so, we first identify the SRs in the DPA \textcolor{black}{statement}. Subsequently, we use the SR \textit{action} to generate the predicate and the remaining SRs to generate the arguments. Recall from Section~\ref{sec:introduction} that a statement in a DPA corresponds to a sentence.

In this step, we design our own method for identifying SRs instead of directly using the results of available NLP tools for semantic role labeling. 
First, the NLP tools have been trained on a large body of generic text. This can result in missing some of the SRs in our list which are pertinent to compliance checking against GDPR, e.g., \textit{reference} and \textit{situation}. Second, a typical NLP semantic role labeler would generate a predicate for each verb in a given statement. In contrast, our representation is based on generating one predicate for each statement in DPA, i.e., extracting one SR \textit{action} to which other SRs in the statement should be related. Generating one predicate is essential for the purpose of our analysis since the SF-based representations of GDPR requirements are centered around one predicate. Our goal in this step is to decompose each statement in the DPA into meaningful phrases that are labeled with similar SRs as in the requirements.  
In addition, comparing multiple predicates for each statement in the DPA against one predicate of each requirement is far too time consuming as reported in Section~\ref{subsec:rqs}. Further, this would lead to multiple answers, most of them being irrelevant, as explained in Section~\ref{subsec:evalProc}. 
To empirically assess the above limitations with respect to the performance of our approach, we define a baseline that works the same way as \textit{DERECHA} with the exception that it employs off-the-shelf NLP tools for generating the SF-based representations in this step. Further details are provided about this baseline in Section~\ref{subsec:evalProc}.

Generating an SF-based representation in \textit{DERECHA} involves two sub-steps, as explained below.

\begin{table*}[t]
\caption{\textcolor{black}{Extraction Rules of Semantic Roles.}}\label{tab:extrules}
 \footnotesize
  \centering
	  \begin{threeparttable}[t]
\begin{tabularx}{\textwidth} {@{} 
  p{0.07\textwidth} |X} 
\toprule
SR  & Description \textbf{(D)}\tnote{$\dag$} and Example \textbf{(E)} from Fig.~\ref{fig:exampledpa} \\
\midrule
\textit{Action} & \textbf{(D)} The \textit{action} is the VP that contains the root verb retrieved from the dependency parsing tree.  The action is the main verb in a statement. We note that a parsing tree contains only one root node, and this is inline with our goal of identifying one predicate for each statement. \textbf{(E)} The \textit{action} in S6 is ``processes''.\\
\midrule
\textit{Actor} & \textbf{(D)} The \textit{actor} is the NP containing the subject of the root verb in the statement (i.e., the \textit{action}). \textbf{(E)} The \textit{actor} in S6 is ``Levico Accounting GmbH''\\
\midrule
\textit{Object} & \textbf{(D)} The \textit{object} is either the NP that contains the object of the root verb \textit{AND} the \textit{action} does not contain any \textit{beneficiary} marker; \textit{OR} the VP that starts with a preposition and is associated with the root verb.  
Our rule accounts for cases where the root verb is directly followed by a PP (e.g., ``assist with''). To differentiate actions that evoke the SR \textit{object} (e.g., ``process'') from other actions that evoke \textit{beneficiary} (e.g., ``inform''), we define markers. \textbf{(E)} The \textit{object} in S6 is ``Company personal data''\\
\midrule
\textit{Beneficiary} & \textbf{(D)} Similar to \textit{object}, the \textit{beneficiary}  is either the NP that contains a preposition linked to the root verb \textit{AND} the \textit{action} contains a \textit{beneficiary} marker, \textit{OR} the NP that contains a subject of a verb that is different from the root verb.  Since the \textit{beneficiary} is an entity that benefits from the \textit{action}, it can be associated directly with the action given that the \textit{action} has a marker that pinpoints the necessity of a \textit{beneficiary}. Alternatively, the \textit{beneficiary} can be an entity associated with another action in the statement. \textbf{(E)} The \textit{beneficiary} in S11 is ``Company''\\
\midrule
\textit{Condition}\tnote{$\ddag$} & \textbf{(D)} The \textit{condition} is the PP, \textit{OR} ADVP, \textit{OR} any phrase  annotated as subordinate clause by the dependency parser given that it contains a \textit{condition} marker. \textbf{(E)} The \textit{condition} in S12 is ``if Company terminates the agreement''. \\
\midrule
\textit{Constraint}\tnote{$\ddag$} & \textbf{(D)} The \textit{constraint} is identified in similar manner as \textit{condition} except that the phrase should contain a \textit{constraint} marker. The \textit{constraint} in S11 is ``without undue delay''\\
\midrule
\textit{Time}\tnote{$\ddag$} & \textbf{(D)} The \textit{time} is the NP, \textit{OR} PP, \textit{OR} ADVP, \textit{OR} any phrase annotated as conjunctive preposition by the dependency parser given that it contains a \textit{time} marker. \textbf{(E)} The \textit{time} in S12 is ``remains effective so long as Levico provides sevices''\\
\midrule
\textit{Reason}\tnote{$\ddag$} & \textbf{(D)} \textit{The reason} is either the VP containing an auxiliary and is linked to a verb other than the root, \textit{OR} the VP containing an open clausal complement and is linked to the root verb. \textbf{(E)} The \textit{reason} in S6 is ``for providing the service of payroll administration''.\\
\midrule
\textit{Situation}\tnote{$\ddag$} & \textbf{(D)} The \textit{situation} is the NP \textit{OR} \textit{VP} that contains a \textit{situation} marker.  \textbf{(E)} The \textit{situation} in S13 is ``termination of any services''.\\
\midrule
\textit{Reference}\tnote{$\ddag$} & \textbf{(D)} The \textit{reference} is the NP that  contains a \textit{reference} marker.  \textbf{(E)} The \textit{reference} in S10 is ``Article 32(1)'' and ``GDPR''.\\
\bottomrule
\end{tabularx}
\begin{tablenotes}
    \item[$\dag$] VP: verb phrase,  NP: noun phrase, PP: prepositional phrase,
    ADVP: adverbial phrase.
    \item[$\ddag$] \textcolor{black}{The extraction rules are adapted from the RE literature~\cite{Sleimi:18,Bhatia:19}.}
\end{tablenotes}
    \end{threeparttable}
\end{table*}

\sectopic{C1) SRs identification.} To identify SRs in  a given statement, we utilize the NLP annotations from step~A (Section~\ref{subsec:stepb}). In particular, we apply a set of rules (listed in Table~\ref{tab:extrules}) over the syntactic information of the statement. 
We rely mostly on the grammatical relations produced by the dependency parser. For instance, the \textit{action} which will be regarded as the predicate is the main verb in the statement. Consequently, the \textit{actor} should be associated with the action as its subject.  We note that any statement without a root verb (e.g., the title of a section) will not have a predicate (i.e., action) and is thus not represented by our method.

Our rules incorporate a set of markers that help identify the SRs. Examples of markers are shown in Table~\ref{tab:markersexamples}. Except for \textit{situation} and \textit{reference}, the markers we use in our work originate from the RE  literature~\cite{Sleimi:18,Bhatia:19}. We improved these markers in two ways. First, we included the nouns and/or verbs that occur in the respective SRs according to our SF-based representations defined for GDPR requirements in step~A. The intuition is to cover what is expected by the GDPR requirements for compliance. Second, we enrich the resulting markers (including the ones reported in the literature and the ones we adapted from GDPR) by including their synonyms from WordNet and VerbNet. 
This way, our markers will likely cover the different wordings applied in the DPA text.

\begin{table}[t]
\caption{Example Markers for Identifying Semantic Roles.}\label{tab:markersexamples}
 \footnotesize
  \centering
	  \begin{threeparttable}[t]
\begin{tabularx}{0.48\textwidth} {@{} 
  p{0.07\textwidth} |X} 
\toprule
SR  & Markers \\
\toprule
\textit{Beneficiary} & inform, report, assist, help, aid, support, remain, notify, provide, give, supply \\
\midrule
\textit{Condition} & if, once, in case, where, when \\
\midrule
\textit{Constraint} & without, on, in accordance with, according to, along, by, under, unless \\
\midrule
\textit{Time} & after, later, prior, before,   earlier, as long as, as soon as\\
\midrule
\textit{Situation} & access, addition, destruction, loss, disclosure,  adherence, modification, termination, expiration \\
\midrule
\textit{Reference} & gdpr, dpa, law, jurisprudence, legislation, agreement, article, contract \\
\bottomrule
\end{tabularx}
    \end{threeparttable}
    \vspace{-1em}
\end{table}

\sectopic{C2) Text spans demarcation.} Once an SR is identified, we demarcate the text span to which this SR should be assigned. Specifically, we use the NLP annotations produced by the text chunking to find the entire phrase where the SR is located. For instance, the SR \textit{action} is assigned to the verb ``employ'' in the statement displayed in Table~\ref{tab:sf-eg}. 
We then find the verb phrase in the chunking results which contains ``employ'', i.e., ``can employ''. This way, we ensure capturing the semantics of the statement in a proper way. 
Except for \textit{action} which has to be identified in a statement only once, other SRs can be identified in the statement multiple times. In this case, we group the $n$ text spans under that SR to get the argument [span$_1$ $| \ldots |$ span$_n$]$_\textit{role}$. For example, the argument [after $|$ end]$_\textit{time}$ contains two spans related to \textit{time}.

As a result of the sub-steps \textbf{\textit{C1}} and \textbf{\textit{C2}} explained above, each statement in the input DPA is sliced into a set of phrases, each one with an SR label. These phrases constitute our SF-based representation for the statement which is further processed in the next step. 

\subsection{Step~D: Enriching Text of Input DPA} \label{subsec:stepd}
In this step, we enrich the DPA text by extending the text spans demarcated in the previous step with semantically-related words. The rationale is to increase the likelihood of finding an overlap when comparing the DPA text against GDPR requirements.
We retrieve semantically-related words from the annotations of the semantic parsing module in Section~\ref{subsec:stepb}. 
We extract from WordNet synonyms for each word in a text span labeled with any SR in the statement. For example, an \textit{object} spanning ``intended changes'' will be enriched with the related words ``planned, alteration, modification''. 
To enrich the text span of \textit{action}, in addition to WordNet, we use the verbs provided in VerbNet under the same class, e.g., the \textit{action} spanning ``can employ'' will be enriched from VerbNet with the verbs ``engage'' and ``hire'', since they are grouped under the same class. \textcolor{black}{To select the synonyms of the words in this step, we opted for the most frequent sense (MFS) of the word. While disambiguating the words can be beneficial, it is outside the scope of our paper. We note that MFS has been widely used as a baseline in the NLP literature and has often shown sufficient performance~\cite{Navigli2007}.} 

In this step, we further replace the actual names of  the controller and processor with the generic place holders ``the controller'' and ``the processor'', e.g., \textit{Levico Accounting GmbH} in the example in Fig.~\ref{fig:exampledpa} is replaced with ``the processor''. As explained in Section~\ref{sec:introduction}, we assume in our work that the name(s) of the controller and processor are given as input by the user. Replacing those named entities helps normalize the text and improve the overall compliance checking.   

\subsection{Step~E: Compliance Checking} \label{subsec:stepe}
In step~E, we check the compliance of the input DPA against GDPR. 

\sectopic{Metadata Requirements. } To verify whether the DPA satisfies the metadata requirements, we do the following.  
R1 and R2 are concerned with the identity and contact details of the controller and processor. Given the names as input by the user, we look for statements in the DPA that contain these names. To find the contact details, we create regular expressions that identify entities including phone numbers, email and postal addresses. Then, we check whether these entities are present in the same statements containing the names of the controller and processor. 

For verifying the remaining requirements (R3 -- R6 and R27 -- R29), we apply a method inspired by Lesk algorithm~\cite{Lesk:86}, one of the traditional methods applied in NLP in the context of word sense disambiguation. The original Lesk algorithm compares multiple senses (meanings) of two words for the purpose of identifying their respective meanings. For example, to disambiguate the word ``cone''  in the phrase ``the pine cones'', Lesk algorithm compares the definition of each sense of the word ``pine'' with the definition of each sense of the word ``cone''. 
The algorithm then computes the number of overlapping words between the definitions of each two senses. The senses with the highest overlap are the ones selected to disambiguate the two words. 
We adapt this algorithm to our compliance checking process. Specifically, for each metadata requirement (among R3 -- R6 and R27 -- R29), we look for a statement in the DPA that has overlapping words with the requirement.  
During our manual analysis of the 24 DPAs (discussed in Section~\ref{sec:qualitative}), we observed that DPA statements often apply the same words when expressing the requirements R3 -- R6 and R27 -- R29. Thus, we believe that using Lesk algorithm in this case is sufficient.

\sectopic{Remaining Requirements. } 
To check whether a requirement is satisfied in the DPA,  
we compare the SF-based representation of the requirement against that of each statement in the DPA. 
Table~\ref{tab:sf-eg} shows a simplified example of checking compliance between a DPA statement and requirement R7. 
Our compliance checking method using SFs comprises four steps, as elaborated in Algorithm~\ref{alg:compliance}.  First, we check the predicates in the two representations. Only if the predicates are sufficiently similar, we subsequently match the arguments, and further compute a score indicating the matching degree. Finally, we conclude the compliance decision for the DPA. We explain each step next.

\begin{table*}[t]
\caption{Example of Compliance Checking using SF-based Representations against R7.}\label{tab:sf-eg}
 \vspace*{-.5em}

 \footnotesize
  \centering
  \begin{threeparttable}

  \begin{tabularx}{\textwidth} {@{} 
  p{0.05\textwidth} @{\hskip 0.5em}|  p{0.1\textwidth} @{\hskip 0.5em}| 
  p{0.07\textwidth} @{\hskip 0.5em} | p{0.1\textwidth} @{\hskip 0.5em}| 
  p{0.18\textwidth} @{\hskip 0.5em} | p{0.05\textwidth} @{\hskip 0.5em}| 
  p{0.1\textwidth} @{\hskip 0.5em} | p{0.2\textwidth} @{\hskip 0.5em} |X}
  \toprule
   \multirow{4}{*}{\textbf{GDPR}} &\multicolumn{8}{l}{{\textbf{R7.}} The processor shall not engage a sub-processor without a prior specific or general written authorization of the controller. } \\
    \cmidrule{2-8}

   &\multirow{2}{*}{\textit{actor}} & \multirow{2}{*}{$\langle p \rangle$} & \multirow{2}{*}{\textit{object}} &\multirow{3}{*}{\colorbox{gray!25}{NOT REQUIRED}} & \multicolumn{3}{c}{\textit{\colorbox{red!25}{constraint}}} \\    
   \cmidrule{6-8}
    & & & & &  & \textit{\colorbox{red!25}{time}} &  \\
    \cmidrule{2-4}\cmidrule{6-8}
    &the processor & not \colorbox{green!25}{engage} & a sub-processor & & without & a prior & specific or general written authorization of the controller. \\
    \midrule
    \midrule
    \multirow{4}{*}{\textbf{DPA}*} &\multicolumn{8}{l}{Levico Accounting GmbH can employ sub-contractors for the performance of the service of Levico's obligations. } \\
        \cmidrule{2-8}

    &\multirow{2}{*}{\textit{actor}} & \multirow{2}{*}{$\langle p \rangle$} & \multirow{2}{*}{\textit{object}} & \multirow{2}{*}{\colorbox{gray!25}{\textit{reason}}}  & \multicolumn{4}{c}{\textit{constraint}} \\    
       \cmidrule{6-8}
    & & & & &  & \textit{time} & &  \\
    \cmidrule{2-8}
     &Levico Accounting GmbH \textbf{processor} & can employ \par \textbf{hire \colorbox{green!25}{engage}}& sub-contractors \textbf{sub-processor} & for the performance of Levico's obligations. \par \textbf{processor duty} & \multicolumn{3}{c}{\colorbox{red!25}{MISSING}} \\
     
    \bottomrule     
    \end{tabularx}

    \begin{tablenotes}
        \item[*] Enriched text from step~D of our approach is highlighted in bold.
    \end{tablenotes}
 \end{threeparttable}
\end{table*}

\renewcommand{\algorithmiccomment}[1]{\bgroup\hfill//~#1\egroup}
\setlength{\textfloatsep}{3pt}
\begin{algorithm}[t!]
\caption{\textcolor{blue}{SF-based Compliance Checking of DPA Statements against a requirement $r$}}\label{alg:compliance}
\begin{algorithmic}[1]

\REQUIRE 
$\mathcal{S}$:~Statements in the input DPA. 

\STATE \texttt{score} $\leftarrow$ 0 \COMMENT{\textcolor{blue}{Set matching degree score of r to zero}}
\STATE \texttt{degree$_i$} $\leftarrow$ 0 
\FOR{$s_i \in \mathcal{S}$}
\STATE Let $\langle p_r\rangle$, $\langle p_i\rangle$ be the predicates of the $r$ and $s_i$, respectively.
\STATE Let $\mathcal{A}_r$, $\mathcal{A}_i$ be the respective set of arguments of $r$ and $s_i$, respectively.
    \IF[\textcolor{blue}{Check compliance only if the two predicates match}]{$\langle p_r\rangle \cap \langle p_i\rangle \neq \phi$ \OR \texttt{sim}$(\langle p_r\rangle,\langle p_i\rangle)>\theta_p$} 
            \FOR{each argument $a_{rj}$ in $\mathcal{A}_r$}
                \STATE \texttt{found} $\leftarrow$ 0  \COMMENT{\textcolor{blue}{Look for matching arguments}}
                \STATE let $\ell_{rj}$ be the SR of $a_{rj}$, and $t_{rj}$ be the text span of $a_{rj}$.
                \FOR{each argument $a_{ik}$ in $\mathcal{A}_i$}
                    \STATE let $\ell_{ik}$ be the SR of $a_{ik}$, and $t_{ik}$ be the text span of $a_{ik}$.
                    \STATE $\texttt{enrich}(t_{ik})$ \COMMENT{\textcolor{blue}{Enrich DPA text with synonyms}}
                    \IF{$(\ell_i$ equals $\ell_r) \And ((t_i \cap t_r \neq \phi)$ \OR \texttt{sim}$(t_i,t_r)>\theta_a))$} 
                        \STATE \texttt{found} $\leftarrow$ \texttt{found} + 1 
                    \ENDIF
                \ENDFOR
                \STATE \texttt{degree$_i$} $\leftarrow$ (\texttt{found$_i$}+1) / (\texttt{len}($\mathcal{A}_r$)+1) \COMMENT{\textcolor{blue}{Compute matching degree of $s_i$}}
            \ENDFOR
    \ENDIF
    \IF{\texttt{score} $<$ \texttt{degree$_i$}}
        \STATE \texttt{score} $\leftarrow$ \texttt{degree$_i$}\COMMENT{\textcolor{blue}{Compute the maximum matching degree score for $r$}}
    \ENDIF
\ENDFOR


\IF[\textcolor{blue}{\it Compliance decision}]{\texttt{score} $\neq 0$}
    \STATE $r$ is satisfied 
\ELSE
    \STATE $r$ is violated 
\ENDIF
\end{algorithmic}
\end{algorithm}

\sectopic{E.1) Matching predicates.} 
For matching two predicates, we specifically consider the verbs occurring in these predicates. We then deem the predicates to be \textit{similar} if at least one of the two conditions below is met (line~6 in Algorithm~\ref{alg:compliance}).  

\noindent\textit{Condition~1} returns \textit{true} if the two predicates have some overlapping verbs.

\noindent\textit{Condition~2} returns \textit{true} if the semantic similarity between the verb in the requirement and at least one verb in the \textcolor{black}{statement} is greater than or equal to a threshold $\theta_p$. 
In our work, we apply the similarity metric (\textit{WuP}) proposed by Wu and Palmer~\cite{Wu:94} and widely-applied in NLP applications~\cite{Corley:05,Newman:10,Shwartz:19}. In brief, WuP is a WordNet-based metric that computes the similarity between two words (or concepts) by finding the path length from the first hypernym in the hierarchy that is shared between the synsets of these concepts~\cite{Pedersen:04}. WuP returns a score between 0.0 (dissimilar) and 1.0 (identical). 

To optimize the efficiency of \textit{DERECHA}, only \textcolor{black}{statements} having similar predicates to the requirements will be further checked for compliance. A \textcolor{black}{statement} that meets none of the conditions above is likely discussing an action that is irrelevant to the one required by GDPR. 

\sectopic{E.2) Matching arguments.} 
This is done based on the arguments in the compliance requirement, since they represent required information by GDPR (lines 7 -- 18 in Algorithm~\ref{alg:compliance}). In particular, we align each argument in the requirement to an \textit{equivalent argument} sharing the same SR in the \textcolor{black}{statement} (e.g., \textit{actor} in Table~\ref{tab:sf-eg}). 

Any argument whose SR exists in the \textcolor{black}{statement} but not in GDPR is considered as \textit{not required}. Conversely, an argument is \textit{missing} if its SR is present in GDPR but not in the statement. For example, the argument with the SR \textit{reason} in Table~\ref{tab:sf-eg} is not required by \textbf{R7}, whereas the one related to \textit{constraint} (including \textit{time}) is missing from the statement. For each pair of equivalent arguments, we check for overlap between the two text spans in a similar way to condition~1 and condition~2 above. For condition~2, we compute the Jaro-Winkler distance instead of WuP that we applied for matching predicates. The reason is that Jaro-Winkler distance is often applied for comparing documents' similarities~\cite{cahyono2019comparison,yulianto2021hybrid,sulaiman2021dissimilarity}. Thus, Jaro-Winkler distance is more suitable in the case of matching arguments since longer text sequences are expected unlike the case of predicates. 

Finally, equivalent arguments are deemed to be \textit{matching} if there is any overlap between the text spans or when the similarity between the text spans is greater than a threshold $\theta_a$. The arguments are deemed \textit{not matching} otherwise.

Following the findings, we empirically set $\theta_p$ to 0.9 and $\theta_a$ to 0.7. We note that for $\theta_p$, unlike $\theta_a$, we experimented with high values ($>0.7$). The rationale is that arguments are only checked when the predicates are deemed similar, i.e., the similarity value between the two predicates passes the threshold $\theta_p$. Considering lower threshold values for $\theta_p$ would entail significant execution time for performing more comparisons involving arguments matching.
\textcolor{black}{To avoid 
excessive computations and further ensure a high level of similarity, we set the first threshold ($\theta_p$) to be relatively high. Once the similarity between two predicates exceeds the threshold (i.e., $>$0.9), we match the arguments. The second threshold ($\theta_a$) is set a bit lower since we already have evidence about the similarity of the predicates.}  
 

In the example shown in Table~\ref{tab:sf-eg}, the predicates are matching since both predicates contain the overlapping verb ``engage''. 
For the sake of providing an example, though not needed here as condition~1 is met, for condition~2, we compute WuP between the verb in the predicate of R7 and each verb in the enriched predicate of the \textcolor{black}{statement}.  
This results in WuP scores of 0.4 between ``engage'' and ``hire'', 0.4 between ``engage'' and ``employ'', and 1.0 between ``engage'' and 
``engage''.
In a similar manner, we match the arguments of R7 with the ones in the statement. Given the overlap in the text spans of the \textit{actor} and \textit{object}, they are considered to be matching. The argument related to \textit{reason} in the statement is skipped since it is not part of the requirement. The argument related to \textit{constraint} has no equivalent argument in the statement and is thus considered not matching. 

\sectopic{E.3) Computing matching degree.} \label{subsec:confscores}
\textit{DERECHA} further computes a score for each compliance requirement indicating 
the maximum matching degree of any \textcolor{black}{statement} satisfying that requirement in the DPA (line 23 in Algorithm~\ref{alg:compliance}). 
Any \textcolor{black}{statement} that does not satisfy any of the two predicate matching conditions above directly receives a score of 0: the statement does not match the requirement at all. 
Otherwise, if a \textcolor{black}{statement} satisfies one predicate matching condition, then the matching degree is computed as a fraction where the numerator equals the total number of arguments found to be matching between the statement and the requirement, and the denominator is the total number of arguments expected by the requirement. 
We add one to both the numerator and denominator corresponding to the matching predicate since the predicate is an essential element of the SF. Including the predicates when computing the matching degree accounts for the cases when a statement matches only the predicate but none of the arguments in the requirement. In this case the statement should still receive some low matching degree that is greater than zero. 
For example, \textbf{R7} in Table~\ref{tab:sf-eg} expects a total of four arguments, namely \textit{actor}, \textit{object}, \textit{constraint}, and \textit{time}. The matching degree between the \textcolor{black}{statement} and R7 is 3/5 (i.e., 0.6), since the statement successfully matches the predicate and two expected arguments of the requirement.  

A requirement can be satisfied by multiple \textcolor{black}{statements} in a DPA. In this case, we assign to the requirement the maximum 
matching degree score across statements. 
\textcolor{black}{Statements} with low matching degrees are due to the lack of matching arguments. However, since these \textcolor{black}{statements} have similar predicates to a requirement, they can still be relevant. 
For such \textcolor{black}{statements}, we therefore consider the matching degree as an indicator of the confidence level at which \textit{DERECHA} predicts that any requirement is satisfied or not in the DPA. This can guide the analyst in verifying and possibly correcting decisions when the confidence level is low. We elaborate on this in Section~\ref{subsec:rqs}.

\begin{figure} 
\includegraphics[width=0.48\textwidth]{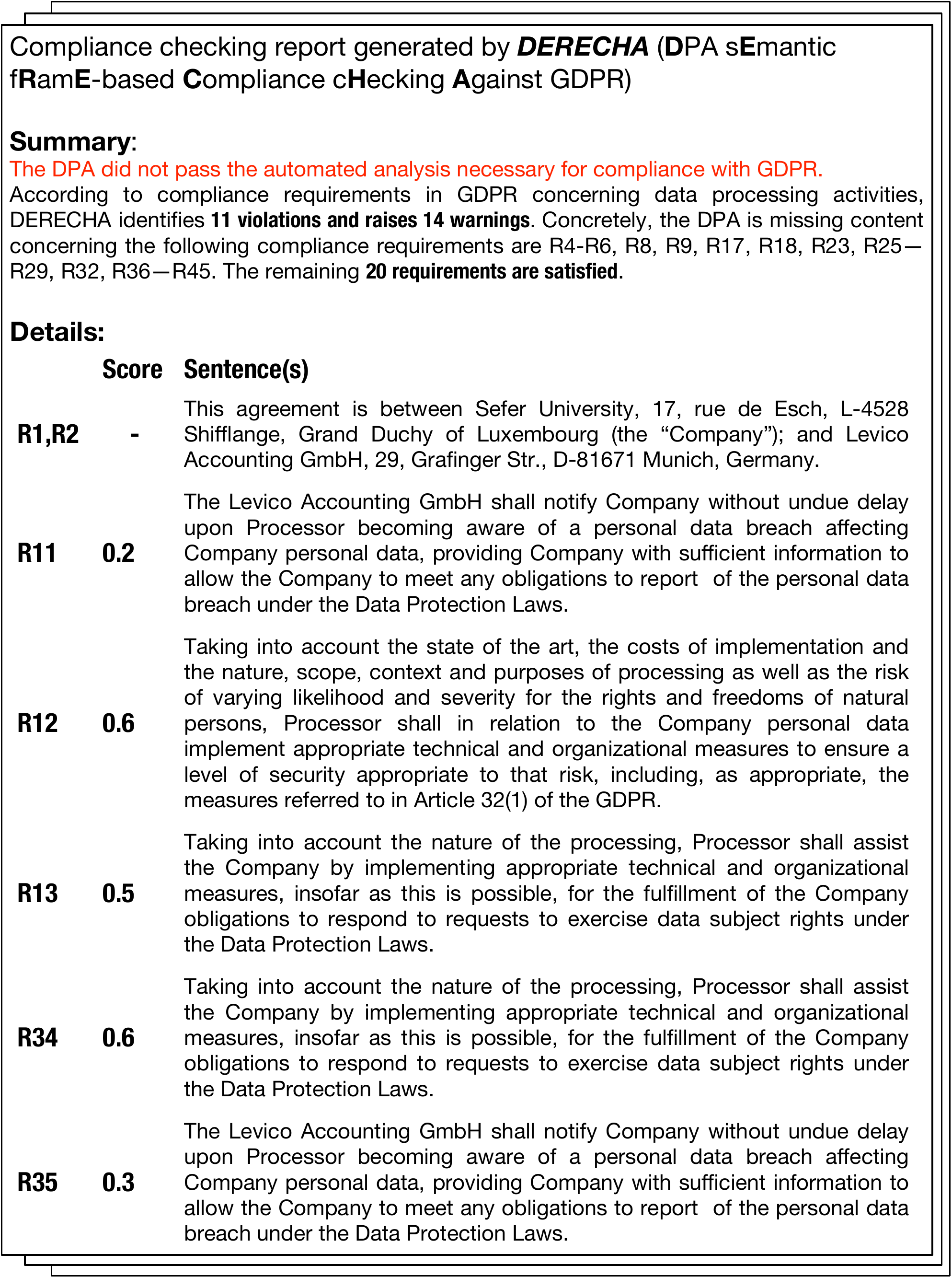}
\caption{\textcolor{black}{Example report generated by \textit{DERECHA}.}}
\label{fig:report}
\vspace{.7em}
\end{figure}

\sectopic{E.4) Compliance decision.} Finally, we make the compliance decision at the DPA level. A requirement is marked as satisfied in the DPA if there is at least one \textcolor{black}{statement} satisfying the requirement. 
Otherwise, the requirement is marked as violated.
\textit{DERECHA} recommends that the DPA should be considered as \textit{not compliant} when at least one mandatory requirement is violated. When optional requirements are violated, \textit{DERECHA} raises a warning to draw the attention of the analyst to missing common practices.
Fig.~\ref{fig:report} shows the detailed report corresponding to the DPA example in Fig.~\ref{fig:exampledpa} as generated by \textit{DERECHA}. 

\section{Evaluation} \label{sec:evaluation}

This section presents the empirical evaluation for our compliance checking approach. 

\subsection{Implementation}
We have implemented our approach in Java. 
We extract the textual content of a DPA (provided as MS Word document) using Apache POI 3.17~\cite{poi}. 
For operationalizing the NLP pipeline applied in step~B of Fig.~\ref{fig:approach}, we use the DKPro 1.10 toolkit~\cite{castilho:14}. In particular, our operationalization employs the OpenNLP tokenizer, POS-tagger and text chunker~\cite{opennlp}, Mate lemmatizer~\cite{mate} and Stanford dependency parser~\cite{stanford}.   
For enriching the textual content of DPA statements (step~D), we rely on the semantic parsing results produced by the NLP pipeline. In particular, we apply extJWNL 1.2~\cite{extJWNL} for accessing WordNet, and JVerbNet 1.2~\cite{JVerbnet} for accessing VerbNet. 
\textcolor{black}{To compute semantic similarity needed in step~E in our approach, we use the \textit{Jaro-Winkler} distance~\cite{jaro1989advances} and the \textit{wup} metric~\cite{Wu:94} as implemented by the WS4J 1.0 library~\cite{ws4j}.}

\subsection{Data Collection Procedure}\label{subsec:docCollection}

To conduct this study, we have collected a total of 54 DPAs \textcolor{black}{  covering diverse sectors and services such as telecommunication, banking, healthcare, postal services, data analytics, cloud and web hosting services.} 
Among them, 34 DPAs have been provided by our industry collaborator (Linklaters LLP) and the rest have been collected from online resources. The goal of our data collection is to manually verify whether a given DPA complies with GDPR requirements and more specifically what statements satisfy which requirements. Our data collection was performed in two phases described next. 

In the first phase, three authors of this paper manually analyzed 24 DPAs for compliance with GDPR. 
This analysis spanned two working weeks. \textcolor{black}{We started by independently analyzing five
DPAs. Specifically, we checked each sentence in the DPAs against the 45 compliance requirements defined in Section~\ref{sec:qualitative} and further labeled the sentence with the requirement(s) it satisfied. 
We then met and discussed our findings and observations in several joint sessions. Once we reached an agreement, we split the remaining DPAs such that each DPA was analyzed by two authors.} 
Finally, we carefully discussed with our collaborating experts the statements that were not straightforward. Our interactions with the legal experts were divided into four sessions over the course of one month, where each session lasted two hours. During this phase, we acquired knowledge about the compliance checking of DPAs and devised annotation guidelines for the second phase.

\begin{table}[!t]
\centering
\caption{\textcolor{black}{Document Collection Results.}}
 \begin{threeparttable}[t]
\begin{tabularx}{0.49\textwidth}{@{}l *{7}{>{\centering\arraybackslash}X}@{}}
\toprule
&  & \multicolumn{2}{c}{\textbf{Total}} & \multicolumn{2}{c}{$Dev$ Data} & \multicolumn{2}{c}{Test Data ($Ev$)} \\
\cmidrule(lr){3-4}\cmidrule(lr){5-6}\cmidrule(lr){7-8}

\textbf{Cat\tnote{$\dag$}} & \textbf{ID} & $\mathcal{D}$ & $\mathcal{S}$ & $\mathcal{D}$ & $\mathcal{S}$ & $\mathcal{D}$  & $\mathcal{S}$  \\
\midrule
\textbf{MD1}	&	R1	&	22  &	29	&	7   &	10	&	15  &	19	\\
\textbf{MD2}	&	R2	&	40  &	93	&	13  &	50	&	27  &	43	\\
\textbf{MD3}	&	R3	&	33  &	45	&	15  &	22	&	18  &	23	\\
\textbf{MD4}	&	R4	&	38  &	133	&	11  &	58	&	27  &	75	\\
\textbf{MD5}	&	R5	&	38  &	270	&	11  &	118	&	27  &	152	\\
\textbf{MD6}	&	R6	&	36  &	126	&	10  &	31	&	26  &	95	\\
MD7	&	R27	&	31  &	914	&	9  &	391	&	22  &	523	\\
MD8	&	R28	&	21  &	99	&	5  &	22	&	16  &	77	\\
MD9	&	R29	&	8  &	9	&	5  &	5	&	3  &	4	\\
\midrule												
\textbf{PO1}	&	R7	&	34  &	62	&	16  &	24	&	18  &	38	\\
\textbf{PO2}	&	R8	&	34  &	52	&	17  &	31	&	17  &	21	\\
\textbf{PO3}	&	R9	&	43  &	83	&	14  &	30	&	29  &	53	\\
\textbf{PO4}	&	R10	&	27  &	55	&	12  &	24	&	15  &	31	\\
\textbf{PO5}	&	R11	&	44  &	64	&	18  &	29	&	26  &	35	\\
\textbf{PO6}	&	R12	&	36  &	206	&	14  &	49	&	22  &	157	\\
\textbf{PO7}	&	R13	&	37  &	85	&	14  &	50	&	23  &	35	\\
\textbf{PO8}	&	R14	&	15  &	22	&	12  &	17	&	3  &	5	\\
\textbf{PO9}	&	R15	&	18  &	25	&	11  &	14	&	3  &	3	\\
\textbf{PO10}	&	R16	&	20  &	27	&	15  &	16	&	7  &	9	\\
\textbf{PO11}	&	R17	&	32  &	34	&	11  &	22	&	5  &	5	\\
\textbf{PO12}	&	R18	&	15  &	17	&	12  &	12	&	21  &	22	\\
\textbf{PO13}	&	R19	&	39  &	67	&	13  &	28	&	26  &	39	\\
\textbf{PO14}	&	R20	&	23  &	25	&	8  &	9	&	15  &	16	\\
\textbf{PO15}	&	R21	&	29  &	48	&	17  &	29	&	12  &	19	\\
\textbf{PO16}	&	R22	&	35  &	49	&	14  &	15	&	21  &	34	\\
\textbf{PO17}	&	R23	&	38  &	59	&	16  &	29	&	22  &	30	\\
\textbf{PO18}	&	R24	&	31  &	53	&	14  &	27	&	17  &	26	\\
\textbf{PO19}	&	R25	&	28  &	36	&	13  &	17	&	15  &	19	\\
PO20	&	R30	&	26  &	29	&	14  &	17	&	12  &	12	\\
PO21	&	R31	&	15  &	19	&	7  &	8	&	8  &	11	\\
PO22	&	R32	&	24  &	35	&	12  &	21	&	12  &	14	\\
PO23	&	R33	&	17  &	23	&	9  &	15	&	8  &	8	\\
PO24	&	R34	&	37  &	67	&	10  &	25	&	27  &	42	\\
PO25	&	R35	&	18  &	25	&	6  &	8	&	12  &	17	\\
\midrule												
\textbf{CR1}	&	R26	&	24  &	34	&	13  &	22	&	11  &	12	\\
CR2	&	R44	&	4  &	5	&	3  &	3	&	1  &	2	\\
CR3	&	R45	&	8  &	21	&	5  &	16	&	3  &	5	\\

\midrule												
CO1	&	R36	&	1  &	1	&	0  &	0	&	1  &	1	\\
CO2	&	R37	&	2  &	3	&	2  &	3	&	0  &	0	\\
CO3	&	R38	&	1  &	1	&	0  &	0	&	1  &	1	\\
CO4	&	R39	&	1  &	1	&	0  &	0	&	1  &	1	\\
CO5	&	R40	&	1  &	1	&	1  &	1	&	0  &	0	\\
CO6	&	R41	&	1  &	1	&	1  &	1	&	0  &	0	\\
CO7	&	R42	&	1  &	1	&	0  &	0	&	1  &	1	\\
CO8	&	R43	&	9  &	14	&	5  &	7	&	4  &	7	\\
\bottomrule
 \end{tabularx}
 \begin{tablenotes}
     \item[$\dag$] Cat: Requirement's category; mandatory requirements are in \textbf{bold}.
   \end{tablenotes}
 \end{threeparttable}
\label{tab:annotations}
\vspace{1em}
\end{table}

The second phase had three third-party annotators read through and examine the remaining \textcolor{black}{30} DPAs for compliance against GDPR. All of the annotators are law students and went through a half-day training on GDPR compliance. We shared with the annotators the guidelines from the first phase, the DPAs to be analyzed and the list of compliance requirements. The annotators were instructed to examine each \textcolor{black}{statement} in the DPA and select all requirements which they deem the statement to satisfy. 
The annotators produced their annotations over a one-month span, during which they declared an average of 40 working hours. To mitigate fatigue, the annotators were recommended to limit their periods of work to a maximum of two hours per day.

To measure the quality of the annotations in this phase, we computed the interrater agreement using Cohen’s Kappa ($\kappa$)~\cite{Cohen:60} on five of the 30 DPAs. 
The average $\kappa$ value across pair-wise agreements of the annotations, split by categories \textit{metadata}, \textit{processor's obligations}, \textit{controller's obligations}, and \textit{controller's rights}, are 0.77, 0.72, 0.82, and 0.79, respectively. $\kappa$ value for the category \textit{controller's obligations} is ``almost perfect agreement''~\cite{Viera:05}, whereas the rest of the three categories indicate ``moderate agreement'' among the annotators. The disagreements were discussed and resolved by the annotators in a subsequent session. 

The overall document collection resulted in analyzing a total of 7048 statements in both phases, out of which 1742 ($\approx$25\%) are marked as satisfying at least one requirement. The DPAs annotated during the first phase were the basis for developing our approach (thereafter referred to as $Dev$ dataset), whereas we used the DPAs annotated by the third-party annotators (thereafter referred to as $Ev$) for addressing our research questions (RQs) and evaluating the performance our approach. Table~\ref{tab:annotations} shows the results of our document collection considering the entire dataset, $Dev$ and $Ev$. The compliance requirements are sorted by category (with mandatory requirements in boldface) and, for each requirement, the table reports the total number of DPAs ($\mathcal{D}$) in which the requirement was found to be satisfied across the 54 DPAs and the total number of statements ($\mathcal{S}$) that satisfy the requirement. 
For example, R9 describes the processor's obligation (\textbf{PO3}) concerning processing personal data only on documented instructions from the controller. In our dataset, R9 was satisfied in 43 DPAs (out of 54 DPAs), and a total of 83 statements were used across the DPAs to comply with R9.   

We observe from the table that mandatory requirements are often satisfied by DPAs while the optional ones are not. According to our collaborating experts, requirements concerning controller's obligations are not necessary for compliance, since GDPR provisions related to DPA focus mostly on the metadata requirements and processor's obligations. We note that, as stated earlier, despite our extensive data collection in terms of DPAs, Table~\ref{tab:annotations} confirms that the small number of statements we have, for each requirement, prevents us from developing \textit{DERECHA} based on machine learning.

\subsection{Evaluation Procedure}
\label{subsec:evalProc}
To answer \textbf{RQ3} and \textbf{RQ4}, we conduct the experiments explained below. 

\sectopic{EXPI.} This experiment addresses \textbf{RQ3}. 
EXPI evaluates whether our approach accurately identifies the violations of compliance requirements in a given DPA. Specifically, we compare the results per requirement in each DPA against the manual annotations in $Ev$. 
We define for each requirement (\textit{R}$_i$) a \textit{true positive (TP)} when \textit{R}$_i$ is violated in the DPA and correctly marked as violated by \textit{DERECHA}, a \textit{false positive (FP)} when \textit{R}$_i$ is satisfied (i.e., not violated) in the DPA, but falsely marked \textit{R}$_i$ as violated, a \textit{false negative (FN)} when \textit{R}$_i$ is violated in the DPA but falsely marked as satisfied, and finally a  \textit{true negative (TN)} when \textit{R}$_i$ is satisfied in the DPA and correctly marked as such. 
Following this, we report the \textit{accuracy (A)}, \textit{precision (P)}, and \textit{recall (R)}, computed as \textit{A~=~(TP+TN)/(TP+FP+FN+TN)}, \textit{P~=~TP/(TP+FP)}, and \textit{R~=~TP/(TP+FN)}, respectively. 
\textcolor{black}{In addition, we report the \textit{F-score} that weighs either precision or recall more highly, computed as (1+$\beta^2$)*(P*R)/($\beta^2$*P+R). In EXPI, we choose to report  F$_{0.5}$, where $\beta = \frac{1}{2}$, indicating that precision is more important than recall in the context of our study as we explain later in this section.}

\sectopic{Baseline.} In EXPI, we further compare our approach against a baseline (referred to as \textbf{B1}), which utilizes off-the-shelf NLP tools for producing SF-Based representations for both the compliance requirements and the DPA statements. 
More specifically, \textbf{B1} applies existing NLP tools for extracting the SRs and generating the SF-based representations. The definitions of the SF-based representations for compliance requirements 
are not applicable for the baseline. The reason is that checking compliance
requires aligning SF-based representations which are generated over the same set of SRs.   

\textcolor{black}{In B1, we apply the semantic role labeling module provided by AllenNLP~\cite{Shi:19} which is based on the BERT language model~\cite{Devlin:18}. 
Since NLP tools generate a predicate-argument structure for each verb in the sentence, \textbf{B1} generates automatically multiple SF-based representations for each statement in the DPA. 
\textbf{B1} automatically generates as well the corresponding SF-based representations for each compliance requirement instead of applying our pre-defined representations. As a result, \textbf{B1} checks the DPA compliance by comparing all SF-based representations generated by AllenNLP for a statement against those generated for a compliance requirement. }
For example, for the requirement ``The processor shall not engage a sub-processor without a prior specific or general written authorization of the controller.'' (R7 in Table~\ref{tab:mandatory-requirements}), \textbf{B1} generates two SF-based representations. The first one contains $\langle \text{\textbf{engage}}\rangle$, $\mathcal{A}=\{$[the processor]$_\textit{argument0}$, [a sub-processor]$_\textit{argument1}\}$, and the second one is $\langle \text{\textbf{written}}\rangle$, $\mathcal{A}=\{$[authorization]$_\textit{argument1}\}$. \textbf{B1} further enriches the DPA text similar to \textit{DERECHA}. 
However, \textbf{B1} limits matching predicates and aligned arguments to exact overlapping text due to efficiency concerns. 
Finally, \textbf{B1} computes the proportion of predicates in a statement that are matching the predicates of the requirement. If this proportion is greater than a certain threshold $\theta_{B1}$, then \textbf{B1} concludes that the statement satisfies the requirement. In our experiments, we empirically tuned $\theta_{B1}$ to 0.30.
For the reasons explained in Section~\ref{subsec:stepa}, SF-based representation does not apply for metadata requirements. Thus, \textbf{B1} is applied only for the three other requirements categories.

\sectopic{EXPII.} This experiment addresses \textbf{RQ4}. We run our approach on a laptop with a 2.3 GHz CPU and 32GB of memory. Our goal is to assess whether execution times suggest that \textit{DERECHA} is applicable in practice.

\subsection{Results and Discussion}\label{subsec:rqs}

\sectopic{RQ3: How accurately can we check the compliance of a given DPA?}

\begin{table*}[!t]
\centering
\caption{Results of Compliance Checking.} 
 \begin{threeparttable}[t]
\begin{tabularx}{0.99\textwidth}{@{}ll *{16}{>{\centering\arraybackslash}X}@{}}
\toprule
 &  & \multicolumn{8}{c}{\textit{DERECHA}}  &  \multicolumn{8}{c}{\textbf{B1}}\\
\cmidrule(lr){3-10}\cmidrule(lr){11-18}
ID  &  \textbf{Cat}\tnote{$\dag$}  &  \textbf{TPs}  &  \textbf{FPs}  &  \textbf{FNs}  &  \textbf{TNs}  &  \textbf{A}  &  \textbf{P}  &  \textbf{R}  &  \textcolor{black}{\textbf{F$_{0.5}$}}
 &  \textbf{TPs}  &  \textbf{FPs}  &  \textbf{FNs}  &  \textbf{TNs}   &  \textbf{A}  &  \textbf{P}  &  \textbf{R}   & \textcolor{black}{ \textbf{F$_{0.5}$}} \\
\midrule
\textcolor{black}{R1} &  \textbf{MD1} & 15 & 3 & 0 & 12 & 90.0 & 83.3 & 100 & 86.2 & - & - & - & - & - & - & - & -\\
R2 &  \textbf{MD2} & 3 & 0 & 0 & 27 & 100 & 100 & 100 & 100 & - & - & - & - & - & - & - & -\\
R3 &  \textbf{MD3} & 5 & 0 & 7 & 18 & 76.7 & 100 & 41.7 & 78.1 & - & - & - & - & - & - & - & -\\
R4 &  \textbf{MD4} & 1 & 0 & 2 & 27 & 93.3 & 100 & 33.3 & 71.4 & - & - & - & - & - & - & - & -\\
R5 &  \textbf{MD5} & 2 & 0 & 1 & 27 & 96.7 & 100 & 66.7 & 90.9 & - & - & - & - & - & - & - & -\\
R6 &  \textbf{MD6} & 2 & 0 & 2 & 26 & 93.3 & 100 & 50 & 83.3 & - & - & - & - & - & - & - & -\\
\textcolor{black}{R7} &  \textbf{PO1} & 4 & 1 & 8 & 17 & 70.0 & 80.0 & 33.3 & 62.5 & 5 & 3 & 7 & 15 & 66.7 & 62.5 & 41.7 & 56.8\\
\textcolor{black}{R8} &  \textbf{PO2} & 8 & 1 & 5 & 16 & 80.0 & 88.9 & 61.5 & 81.6 & 12 & 16 & 1 & 1 & 43.3 & 42.9 & 92.3 & 48.0\\
\textcolor{black}{R9} &  \textbf{PO3} & 1 & 1 & 0 & 28 & 96.7 & 50.0 & 100 & 55.6 & 0 & 1 & 1 & 28 & 93.3 & 0.0 & 0.0 & 0.0\\
R10 &  \textbf{PO4} & 5 & 2 & 10 & 13 & 60.0 & 71.4 & 33.3 & 58.1 & 10 & 12 & 5 & 3 & 43.3 & 45.5 & 66.7 & 48.6\\
\textcolor{black}{R11} &  \textbf{PO5} & 4 & 1 & 0 & 25 & 96.7 & 80.0 & 100 & 83.3 & 3 & 19 & 1 & 7 & 33.3 & 13.6 & 75.0 & 16.3\\
\textcolor{black}{R12} &  \textbf{PO6} & 2 & 0 & 6 & 22 & 80.0 & 100 & 25.0 & 62.5 & 3 & 8 & 5 & 14 & 56.7 & 27.3 & 37.5 & 28.9\\
\textcolor{black}{R13} &  \textbf{PO7} & 3 & 1 & 4 & 22 & 83.3 & 75.0 & 42.9 & 65.2 & 5 & 22 & 2 & 1 & 20.0 & 18.5 & 71.4 & 21.7\\
\textcolor{black}{R14} &  \textbf{PO8} & 27 & 3 & 0 & 0 & 90.0 & 90.0 & 100 & 91.8 & 3 & 0 & 23 & 4 & 23.3 & 100 & 11.5 & 39.4\\
\textcolor{black}{R15} &  \textbf{PO9} & 17 & 3 & 6 & 4 & 70.0 & 85.0 & 73.9 & 82.5 & 22 & 5 & 1 & 2 & 80.0 & 81.5 & 95.7 & 84.0\\
\textcolor{black}{R16} &  \textbf{P1O} & 18 & 3 & 7 & 2 & 66.7 & 85.7 & 72.0 & 82.6 & 15 & 1 & 10 & 4 & 63.3 & 93.8 & 60.0 & 84.3\\
\textcolor{black}{R17} &  \textbf{PO11} & 3 & 3 & 6 & 18 & 70.0 & 50.0 & 33.3 & 54.5 & 1 & 1 & 8 & 20 & 70.0 & 50.0 & 11.1 & 29.4\\
\textcolor{black}{R18} &  \textbf{PO12} & 27 & 2 & 0 & 1 & 93.3 & 93.1 & 100 & 94.4 & 23 & 2 & 4 & 1 & 80.0 & 92.0 & 85.2 & 90.6\\
\textcolor{black}{R19} &  \textbf{PO13} & 2 & 1 & 2 & 25 & 90.0 & 66.7 & 50.0 & 62.5 & 0 & 2 & 4 & 24 & 80.0 & 0.0 & 0.0 & 0.0\\
\textcolor{black}{R20} &  \textbf{PO14} & 12 & 5 & 3 & 10 & 73.3 & 70.6 & 80.0 & 72.3 & 11 & 12 & 4 & 3 & 46.7 & 47.8 & 73.3 & 51.4\\
\textcolor{black}{R21} &  \textbf{PO15} & 13 & 3 & 5 & 9 & 73.3 & 81.3 & 72.2 & 79.3 & 2 & 0 & 16 & 12 & 46.7 & 100 & 11.1 & 38.4\\
\textcolor{black}{R22} &  \textbf{PO16} & 4 & 1 & 5 & 20 & 80.0 & 80.0 & 44.4 & 69.0 & 8 & 9 & 1 & 12 & 66.7 & 47.1 & 88.9 & 52.0\\
\textcolor{black}{R23} &  \textbf{PO17} & 4 & 3 & 4 & 19 & 76.7 & 57.1 & 50.0 & 55.6 & 4 & 12 & 4 & 10 & 46.7 & 25.0 & 50.0 & 27.8\\
\textcolor{black}{R24} &  \textbf{PO18} & 13 & 3 & 0 & 14 & 90.0 & 81.3 & 100 & 84.4 & 4 & 1 & 9 & 16 & 66.7 & 80.0 & 30.8 & 60.6\\
\textcolor{black}{R25} &  \textbf{PO19} & 10 & 4 & 5 & 11 & 70.0 & 71.4 & 66.7 & 70.4 & 7 & 5 & 9 & 9 & 53.3 & 58.3 & 43.8 & 54.7\\
\textcolor{black}{R26} &  \textbf{CR1} & 16 & 3 & 6 & 5 & 70.0 & 84.2 & 72.7 & 81.6 & 18 & 10 & 1 & 1 & 63.3 & 64.3 & 94.7 & 68.7\\
\midrule
\textcolor{black}{R27} &  MD7 & 6 & 2 & 2 & 20 & 86.7 & 75.0 & 75.0 & 75.0 & - & - & - & - & - & - & - & -\\
\textcolor{black}{R28} &  MD8 & 6 & 0 & 7 & 17 & 76.7 & 100 & 46.2 & 81.1 & - & - & - & - & - & - & - & -\\
\textcolor{black}{R29} &  MD9 & 28 & 0 & 0 & 2 & 100 & 100 & 100 & 100 & - & - & - & - & - & - & - & -\\
\textcolor{black}{R30} &  PO20 & 3 & 3 & 7 & 17 & 76.7 & 50.0 & 30 & 44.1 & 4 & 4 & 6 & 16 & 66.7 & 50.0 & 40.0 & 47.6\\
\textcolor{black}{R31} &  PO21 & 30 & 0 & 0 & 0 & 100 & 100 & 100 & 100 & 29 & 0 & 1 & 0 & 96.7 & 100 & 96.7 & 99.3\\
\textcolor{black}{R32} &  PO22 & 15 & 4 & 3 & 8 & 76.7 & 78.9 & 83.3 & 79.8 & 11 & 8 & 7 & 4 & 50.0 & 57.9 & 61.1 & 58.5\\
\textcolor{black}{R33} &  PO23 & 16 & 6 & 3 & 5 & 70.0 & 72.7 & 84.2 & 74.8 & 3 & 0 & 19 & 8 & 36.7 & 100 & 13.6 & 44.0\\
\textcolor{black}{R34} &  PO24 & 3 & 0 & 0 & 27 & 100 & 100 & 100 & 100 & 2 & 3 & 1 & 24 & 86.7 & 40.0 & 66.7 & 43.5\\
\textcolor{black}{R35} &  PO25 & 12 & 5 & 6 & 7 & 63.3 & 70.6 & 66.7 & 69.8 & 17 & 12 & 1 & 0 & 56.7 & 58.6 & 94.4 & 63.4\\
R36 &  CO1 & 29 & 1 & 0 & 0 & 96.7 & 96.7 & 100 & 97.3 & 14 & 0 & 15 & 1 & 50.0 & 100 & 48.3 & 82.4\\
R37 &  CO2 & 30 & 0 & 0 & 0 & 100 & 100 & 100 & 100 & 26 & 0 & 4 & 0 & 86.7 & 100 & 86.7 & 97.0\\
\textcolor{black}{R38} &  CO3 & 28 & 0 & 1 & 1 & 96.7 & 100 & 96.6 & 99.3 & 25 & 0 & 5 & 0 & 83.3 & 100 & 83.3 & 96.1\\
R39 &  CO4 & 29 & 1 & 0 & 0 & 96.7 & 96.7 & 100 & 97.3 & 13 & 0 & 16 & 1 & 46.7 & 100 & 44.8 & 80.2\\
R40 &  CO5 & 30 & 0 & 0 & 0 & 100 & 100 & 100 & 100 & 29 & 0 & 1 & 0 & 96.7 & 100 & 96.7 & 99.3\\
R41 &  CO6 & 30 & 0 & 0 & 0 & 100 & 100 & 100 & 100 & 13 & 0 & 17 & 0 & 43.3 & 100 & 43.3 & 79.2\\
R42 &  CO7 & 29 & 1 & 0 & 0 & 96.7 & 96.7 & 100 & 97.3 & 26 & 1 & 3 & 0 & 86.7 & 96.3 & 89.7 & 94.9\\
\textcolor{black}{R43} &  CO8 & 20 & 2 & 6 & 2 & 73.3 & 90.9 & 76.9 & 76.2 & 24 & 4 & 2 & 0 & 80.0 & 85.7 & 92.3 & 86.9\\
R44 &  CR2 & 29 & 1 & 0 & 0 & 96.7 & 96.7 & 100 & 97.3 & 24 & 1 & 5 & 0 & 80.0 & 96.0 & 82.8 & 93.0\\
R45 &  CR3 & 24 & 3 & 3 & 0 & 80.0 & 88.9 & 88.9 & 88.9 & 20 & 1 & 7 & 2 & 73.3 & 95.2 & 74.1 & 90.1\\
\midrule
\midrule
\multicolumn{2}{c}{\textcolor{black}{Summary}}	 &   \textbf{618}	 & 	\textbf{76}	 & 	\textbf{132}	 & 	\textbf{524}	 & 	\textcolor{black}{\textbf{84.6}}	 & 	\textcolor{black}{\textbf{89.1}}	 & \textcolor{black}{	\textbf{82.4}}	 & 	\textcolor{black}{\textbf{87.6}}	 & 	\textbf{436} & \textbf{175} & \textbf{226} & \textbf{243} & \textbf{62.9} & \textcolor{black}{\textbf{71.4}} & \textcolor{black}{\textbf{65.9}} & \textcolor{black}{\textbf{70.2}}	\\
\bottomrule
\end{tabularx}
 \begin{tablenotes}
     \item[$\dag$] Cat: Requirement's category; mandatory requirements are in \textbf{bold}.
   \end{tablenotes}
 \end{threeparttable}
 \label{tab:RIresults}
 \vspace*{1em}
\end{table*}



Table~\ref{tab:RIresults} shows the results of EXPI; on the left-hand side, the results of \textit{DERECHA}, and on the right-hand side the
results of \textbf{B1} introduced in Section~\ref{subsec:evalProc}.
As explained in Section~\ref{subsec:docCollection}, the results are obtained by running
\textit{DERECHA} and \textbf{B1} on the evaluation set ($Ev$), which is comprised of 30 DPAs (7048 statements) on which we verify the compliance of our 45 requirements. \textcolor{black}{\textit{DERECHA} correctly finds 618 requirement violations (out of 750) and 524 satisfied requirements, but also leads to 76 false violations}. As pinpointed in section \ref{subsec:evalProc}, \textbf{B1} is applied only to the categories of requirements PO, CO, and CR. Hence, we compare the performance of \textbf{B1} against \textit{DERECHA} considering only these categories. \textcolor{black}{Specifically, \textit{DERECHA} correctly identifies 550 violated requirements (out of 661) and 348 satisfied requirements (out of 419), while introducing 71 false violations. In comparison, \textbf{B1} finds 436 violations, 243 satisfied requirements and introduces 175 false violations. 
To summarize, 182 errors (FPs+FNs) are produced by \textit{DERECHA} for both mandatory and optional requirements in these three categories, whereas \textbf{B1} produces 401 (219 more errors). Overall, \textit{DERECHA} outperforms \textbf{B1} by an average of $\approx$20 percentage points (pp) in accuracy, $\approx$17 pp in precision, recall, and F$_{0.5}$}.


\textcolor{black}{In the context of our study, we favor high precision over high recall. In the case of unsatisfactory recall, the human analyst needs to go through the subset of statements that are marked by \textit{DERECHA} as satisfying a requirement in order to determine whether this is actually the case.   
Unsatisfactory precision, on the other hand, indicates that some predicted violations cannot be trusted with a high level of confidence. In such a case, the human analyst would have no choice but to check all statements in the DPA to decide whether at least one of them actually satisfies a requirement. However, doing such a thing would defeat the very purpose of \textit{DERECHA}.   
In contrast, addressing unsatisfactory recall can be done with relative ease as the number of statements satisfying any requirement usually represent a small percentage. We discuss next the trade-off between additional manual validation and accuracy improvement.  
} 

\begin{figure} [t!]
    \centering
    \includegraphics[width=0.4\textwidth]{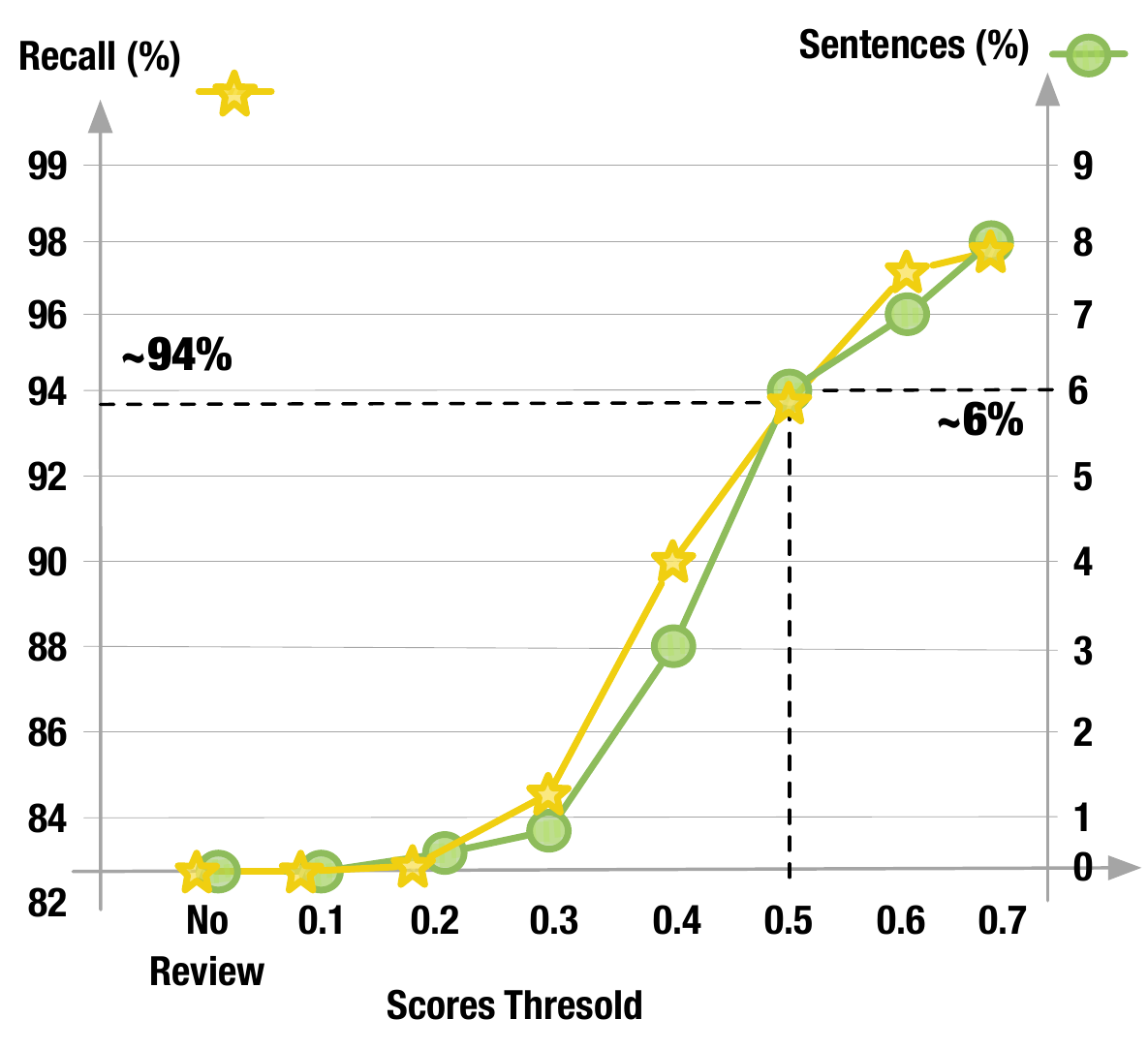}
        \vspace{-.2em}

    \caption{\textcolor{black}{Validation-accuracy trade-off.}}
    \label{fig:confidence}
    \vspace{.5em}
\end{figure}

FNs entail that our approach can mispredict a requirement as being satisfied by at least one \textcolor{black}{statement} in the DPA. We carefully checked the matching degrees (scores defined in Section~\ref{subsec:confscores}) of these FNs
and observed that, in \textcolor{black}{86} out of \textcolor{black}{132} cases, such requirements are assigned matching degree scores $\leq$0.5. 
By reviewing a fraction of \textcolor{black}{statements}, which are predicted to satisfy any requirement with a score below some predefined threshold, the analyst can identify statements that actually do not satisfy any requirement. One possible heuristic that was observed to work well is, for each requirement, to only analyze the statement with maximum score if it is below the selected threshold. 
Assuming we apply the above heuristic, Fig.~\ref{fig:confidence} shows the gain in recall achieved, for instance, when considering a threshold of 0.5: recall can be improved from \textcolor{black}{82.4\% to 93.9\%} at the cost of reviewing $\approx$6\% of the statements. For an average-sized DPA with 200 statements, this percentage corresponds to approximately 12 statements. Further increases in recall are possible but are associated with higher reviewing effort and the right trade-off depends on context. Eliminating FNs through manual reviews also slightly improves precision and accuracy. \textcolor{black}{For example, the manual review mentioned above increases precision to 90.3\% and accuracy to 91.0\%}. 
Selecting a threshold at which the user reviews \textit{DERECHA}'s output (matching statements) in order to improve accuracy depends on the resources available and therefore the number of statements that can realistically be reviewed using our heuristic. 
Users who can afford it might opt for reviewing all of the matching statements with maximum scores for each requirement to achieve a near perfect recall (\textcolor{black}{with $\approx$90.6\% precision and $\approx$93.1\% accuracy}). In our DPAs, doing so would result in reviewing about 9\% of statements on average across DPAs.


\textcolor{black}{\textit{DERECHA} achieves an average precision of 81.5\% and 93.9 \% when identifying violations of  mandatory (R1 -- R26) and optional requirements (R27 -- R45), respectively. In other words, \textit{DERECHA} introduces false violations (FPs) in a total of 76 cases, out of which 50 are related to mandatory requirements}. We analyzed these cases to determine the root causes, which we explain below. We further illustrate each cause with an example from our data collection, and report the number of FPs attributable to that cause. 

\begin{itemize}[wide]
    \item [\textbf{C1.}] \textit{Inconsistent reference to the controller and/or processor:} 22 FPs are caused by this error.
    We observed that in addition to using a concrete named entity representing the controller or the processor (e.g., Levico in Fig.~\ref{fig:exampledpa}), many DPAs use other words to refer to the controller or processor. For example, the following statement refers to the processor as \textit{data importer} and to the controller as \textit{data exporter}: ``The data importer agrees at the request of the data exporter to submit its data processing facilities for inspection of the processing activities covered by the clauses which shall be carried out by the data exporter''. This statement should be predicted as satisfying R22 (see Table~\ref{tab:mandatory-requirements}). However, it is unlikely for \textit{DERECHA} to identify this statement correctly due to the way it is expressed and the additional unconventional names used in place of processor and controller. As an attempt to reduce such errors, we can modify \textit{DERECHA} in a way such that the user of our tool can input multiple names for the controller or the processor accounting for both the named entity (e.g., Levico GmbH) and references (e.g., service provider). The user might be familiar with some commonly used names, e.g., \textit{contractor}, \textit{client}, \textit{customer}, or \textit{operator}.

    \item [\textbf{C2.}] \textit{Loss of context:}
    \textcolor{black}{39 FPs are caused by this error}. 
    In our work, we use the NLP pipeline to automatically demarcate statements in a given DPA, where each statement corresponds to one sentence as produced by the sentence splitter. Statements included in itemized or enumerated lists can suffer from loss of context when the statements cover grammatically incomplete sentences (marked as \textbf{S1} -- \textbf{S5}). For example, the following list spans five statements ``Where Processor uses subprocessors for the processing of personal data on behalf of Controller, Processor shall: \textbf{[S1]} 
    \begin{itemize}
        \item impose the same obligations on the engaged subprocessors; \textbf{[S2]}
        \item transfer only the data that is reasonably necessary for the purposes of providing the services; \textbf{[S3]}
        \item ensure all the time the security of the shared personal data, using encryption strategies; \textbf{[S5]}
        \item remain fully liable to the Controller for the performance of subprocessors.'' \textbf{[S5]}
    \end{itemize} 
    In the above example, the combination of the first two statements (\textbf{S1} and \textbf{S2}) should be predicted as satisfying R23. However, considering these two statements separately leads to falsely predicting a violation of R23 instead.  
    
    \item [\textbf{C3.}] \textit{Noisy text in DPA:} 
    \textcolor{black}{15} FPs are caused by this error. 
    Statements in a given DPA are often verbose, i.e., containing content that is considered irrelevant regarding compliance with requirements. 
    Such cases can result in poor performance in the NLP syntactic parsing on which our SR extraction rules depend. For example, the following statement is long and contains multiple clauses that are hard to accurately parse: ``Controller will permit Processor to take reasonable steps, according to a reasonable notice and during normal business hours, at Processor's cost to assess compliance by Controller with its obligations under this DPA, including by inspecting Controller's data processing facilities, procedures and documentation (limited to a maximum of one inspection in any twelve month period, or such further occasions as may be required by Privacy Law or if there is an actual Personal Data Breach by Controller)''.
    
    Handling errors caused by \textbf{C2} and \textbf{C3} could be achieved with heuristics to identify and paraphrase itemized lists, or filter out noisy content from DPA statements. 
\end{itemize}

Table~\ref{tab:RIresults} further suggests that the baseline \textbf{B1} has two disadvantages. First, compared to our approach, \textbf{B1} checks the compliance of a statement in DPA against a compliance requirement considering multiple verbs (predicates). This not only increases the processing time exponentially but also leads to poor results. Some of these predicates include very common verbs that can be found often in statements such as "mean", "take", and "write". 
Having multiple predicates increases the likelihood of concluding a match between statements and  requirements, consequently leading to more FNs (e.g., R14 and R33). 
Second, \textbf{B1} relies fully on the NLP tool for semantic role labeling (SRL). In some cases, \textbf{B1} could not identify any statement satisfying a given requirement since the SRL tool could not accurately parse the statement. This led to producing more FPs (e.g. R11 and R13). In contrast, \textit{DERECHA} employs a set of rules that are likely to decompose the statement into meaningful phrases and thus conclude more accurately about their compliance against requirements. 
Our analysis shows that off-the-shelf NLP tools are unable to identify SRs in a legal text and are thus not adequate for checking the compliance of DPAs.

\begin{tcolorbox}[arc=0mm,width=\columnwidth,
                  top=0mm,left=0mm,  right=0mm, bottom=0mm,
                  boxrule=1pt,colback=blue!5!white,colframe=black]
\textbf{The answer to RQ3 is that \textit{DERECHA} is \textcolor{black}{84.6\% accurate in identifying both satisfied and violated GDPR requirements in DPAs. Specifically, \textit{DERECHA} identifies violated requirements with a precision of 89.1\% and a recall of 82.4\%.}} Compared to a baseline that relies on existing NLP tools, \textit{DERECHA} provides an accuracy gain of $\approx$20 pp. Our approach also computes matching scores that allow the analyst to review a small percentage of statements with low scores and thus eases the identification of FNs to further increase accuracy. 
\end{tcolorbox}

\sectopic{RQ4: How efficient is our approach in terms of the execution time?}

Recall that our approach consists of five steps as shown in Fig.~\ref{fig:approach}. Step~A is concerned with manually defining SF-based representations for the requirements in GDPR. 
Step~A took about a month, including reviewing related work to define the set of SRs. 

This step is a one-off and performed only occasionally upon changes in the regulation, i.e., new requirements become relevant to DPA compliance. %
For analyzing the complete evaluation set of 30 DPAs, running our approach requires $\approx$90 minutes (corresponding to an average of $\approx$0.77 second for processing a given statement). 
The detailed execution time required by each step of our approach is as follows. Preprocessing (step~B) requires a total of 1 minute ($\approx$0.009 second per statement). Generating the SF-based representations automatically (step~C) is dominated by identifying SRs. This step takes in total about 52 minutes ($\approx$0.44 second per statement). We note that step~C consumes most of the time in our approach since our rules are built over syntax parsing. 
Finally, steps~D~and~E require 37 minutes ($\approx$0.32 second per statement). For efficiency, we implement step~D (text enrichment) as part of step~E (compliance checking), i.e., we perform text enrichment of arguments only when the statements in DPA have passed predicate matching.
Our approach runs $\approx$10 times faster than the baseline \textbf{B1}. The reason is, \textbf{B1} compares all the predicates in a given statement against all the predicates in a requirement leading to a significantly longer computational time.

\vspace*{.5em}
\begin{tcolorbox}[arc=0mm,width=\columnwidth,
                  top=0mm,left=0mm,  right=0mm, bottom=0mm,                  boxrule=1pt,colback=blue!5!white,colframe=black]
To conclude, \textbf{the answer to RQ4 is that our approach has a running time enabling its application in practice.} 
Running our approach on an average-sized DPA (with 200 statements) takes $\approx$\textcolor{black}{2.5} minutes to verify its compliance against all 45 requirements. 
\end{tcolorbox}


\section{Related Work} \label{sec:related}

Regulatory compliance is a long-standing research topic in the RE literature~\cite{Breaux:08,Ghanavati:11,Massey:14,Ghanavati:14,Evans:17,Akhigbe:21,Abualhaija:22,Breaux:22}. While little attention has been given to the compliance of DPAs, recent research explores GDPR compliance of data sharing. 
\textcolor{black}{We therefore discuss current research on data sharing platforms (e.g., cloud service providers) since a GDPR-compliant DPA provides a complete set of compliance requirements that must be addressed in data processing activities.} 
In addition, we address below two other broad topics that are closely related to our proposed approach for checking compliance, namely elicitation of compliance requirements from regulation, and automated support for compliance checking of requirements and software systems. Naturally, we focus our discussions in this section on GDPR.

\subsection{GDPR Compliance of Data Processing Activities}
Several studies analyze the impact of GDPR on data sharing platforms~\cite{horak:19, loideain:19}. Urban et al.~\cite{urban:20} study the behavior of online advertising companies in data tracking and information sharing with respect to the GDPR. Alhazmi and Arachchilage~\cite{alhazmi:21a} investigate the obstacles that developers encounter when implementing the privacy provisions in GDPR, including lack of familiarity with legal text.  In a follow-up work, the authors~\cite{alhazmi:21b} propose a game design framework to help software developers better understand and implement privacy-related requirements in GDPR. Shastri et al.~\cite{shastri:21} introduce anti-patterns, referring to practices of cloud-scale systems that serve their originally intended purpose well but hinder their compliance with GDPR. For example, such systems are developed to store personal data without having a clear timeline for deleting and sharing them with other applications. While this feature is naturally beneficial in terms of revenues from the system, it violates the storage and purpose limitations of GDPR. 

We observe from the above-mentioned work that implementing the GDPR requirements in practice is challenging for developers. Even developers that are familiar with GDPR still struggle with identifying the complete set of requirements that is most relevant to their work. \textcolor{black}{A compliant DPA contains a complete set of legally binding requirements that must be adhered to by both the controller and processor. Recall  our example in Section~\ref{sec:introduction} which describes an accounting office (the processor) that provides payroll administration services to Sefer University (the controller). In this case, the engineers working in the accounting office should specify explicit requirements to demonstrate that their payroll software complies with GDPR. Since the DPA contains statements specific to the context of this service, such requirements can be defined based on the DPA provided, assuming it has been properly verified and found to be GDPR-compliant. 
For example, the engineers should define 
a communication procedure, satisfying time requirements,  in case of personal data breach (pursuant to S11 in Fig.~\ref{fig:exampledpa}), a module that enables rectifying personal data of Sefer University's employees (data subjects) upon request (pursuant to  S9), and an authorization module that allows only authorized users to access personal data (pursuant to S10). Our approach, presented in this paper, provides automated assistance to the developers and requirements engineers to verify the compliance of a DPA prior to relying on it for
specifying compliance requirements for developing GDPR-compliant systems.}


\subsection{Elicitation of Compliance Requirements from Regulation} 
In the RE literature, representing requirements in regulations (including GDPR) relies heavily on model-driven engineering~\cite{Otto:07,Siena:09,Maxwell:09,Ghanavati:09,Ingolfo:14,Soltana:14,Zeni:15,Pullonen:19,VaneziKKPP20,Amaral:21,Amaral:21b}. 
Other methods for representing compliance requirements have however been explored. Ayala-Rivera and Pasquale~\cite{Ayala:18} propose GuideMe, a systematic approach that facilitates the elicitation of requirements linking GDPR data protection obligations with privacy concerns that should be implemented in a software system. In an early work, Breaux et al.~\cite{Breaux:06} extract and prioritize rights and obligations from the US healthcare regulation and represent them by applying semantic parameterization, a method that enables expressing rights and obligations in restricted natural language statements. The authors derive from the regulation constraints and obligations where each phrase is attributed to an element in their representation. Following this, Breaux and Anton’s~\cite{Breaux:07} propose a frame-based requirements analysis method that extracts semi-formal representations of requirements from regulations. More recently, formal languages~\cite{Binsbergen:20, xu2021ontology,mashkoor2021validation, tokas2022static} and predefined templates~\cite{Sangaroonsilp:21} have also been introduced in the literature to represent compliance requirements. 
{\color{black}To our knowledge, the only work to date that addresses the compliance of DPAs against GDPR is the work by Amaral et al.~\cite{Amaral:21}. The authors create a conceptual model that characterizes the DPA-compliance related information content according to GDPR provisions. 

Compared with Amaral et al.'s work, we go one step further and present automated support for verifying the compliance of DPAs against GDPR. 
Another major difference (also compared to the above listed work) is that we define DPA-related compliance requirements as ``shall'' requirements.} 
Our choice is motivated by the familiarity of both requirements engineers and legal experts with this format. Requirements engineers know from the common templates in RE (e.g., Rupp's~\cite{Pohl:11} and EARS~\cite{Mavin:09}) that a modal verb (``shall'' in this case) is an essential element indicating the importance of that requirement. Legal experts are also familiar with this format since ``shall'' is often used in drafting requirements in regulations~\cite{Krapivkina:17}.

The use of SRs (also referred to as semantic metadata in the RE literature) has been also discussed in the context of regulatory compliance. Semantic metadata is a prerequisite for deriving compliance requirements~\cite{Breaux:08,Siena:09,Bhatia:16}. It further facilitates transforming legal text to formal specifications~\cite{Maxwell:10}. Several strands of work define legal concepts pertinent to requirements in regulations as semantic metadata types. For a detailed discussion on semantic metadata in the RE literature, we refer the reader to the work of Sleimi et al.~\cite{Sleimi:18}. In our work, we  refine and adapt a set of 10 SRs from literature~\cite{Zeni:15, boella2016eunomos, ingolfo2014nomos, zeni2016building, ghanavati2014goal, massey2009prioritizing, maxwell2010production}. 

\subsection{Automated Support for Compliance Checking} 

Automated and semi-automated approaches using machine learning (in combination with NLP)~\cite{Tesfay:18,lippi:19, Nejad:20,Sanchez:21,Hamdani:21,Aberkane:21,Amaral:21b}, as well as manual approaches through crowd-sourcing~\cite{Liu:14,Wilson:16,Bhatia:16,linden:20}, have been studied in the RE literature for classifying the textual content according to privacy regulations. With respect to GDPR, automated compliance has been widely studied~\cite{Park2019,Benjumea2019,Verderame2020,Guaman2021,Cozar2022}. Several approaches rely on semantic web as an enabler~\cite{Palmirani:18,Bonatti:20}. Li et al.~\cite{Li:20} identify a set of operationalized GDPR principles and further develop a tool to test these derived GDPR privacy requirements. The authors apply design science and their findings reveal that GDPR can be operationalized and tested through automated means. Nazarenko et al.~\cite{Nazarenko:21} propose enriching legal text with semantic information both at a sentence-level (e.g., annotating that a sentence contains an exception) and at a phrasal-level (e.g., annotating that a phrase represents a legal entity) to facilitate semantic search and querying legal text. The authors apply their proposed approach on the French version of GDPR. Sleimi et al.~\cite{Sleimi:18} propose an automated rule-based method over the NLP syntax parsing to extract legal semantic metadata, e.g., an actor or a sanction, from regulations. They evaluate their approach on traffic law. Bhatia et al.~\cite{Bhatia:19} propose using semantic frames to detect incompleteness in privacy policies. Specifically, the authors manually analyze 15 privacy policies and conclude a set of 17 semantic roles that are expected to be present in other policies from the same domain. The authors use these SRs to further study the effect of an incomplete privacy statement (i.e., a statement that misses semantic roles) on the user's perspective of a privacy risk.

The approaches of Sleimi et al.~\cite{Sleimi:18} and Bhatia et al.~\cite{Bhatia:19} are the closest to our work. In contrast, our work utilizes automatically-generated semantic frames for enabling automated compliance checking of DPAs at a phrasal level.

\textcolor{black}{Another strand of research focuses on checking the compliance of websites and mobile applications against applicable laws. Extensive analyses of popular websites show that the variance in how organizations interpret privacy and data protection standards can lead to legal documents (e.g., privacy policies) which are not informative for individuals~\cite{chen:20,degeling2019}. Relevant to this, the lack of templates for drafting such legal documents can cause discrepancies between the data processing activities and what is actually disclosed to individuals~\cite{rahman2022}, e.g., about the purpose of data processing (an essential requirement in GDPR)~\cite{aberkane2022}. Several approaches are proposed to check the compliance of websites by classifying the content of their disclosed privacy policies according to the provisions of GDPR. Our work addresses three main challenges introduced in the above-listed work. First, we derive the compliance requirements from GDPR in collaboration with legal experts, which alleviates the risk of misinterpreting GDPR. We further make these requirements publicly available. Second, our work addresses the compliance of DPAs, another essential source for requirements for data processing activities in software systems. Our proposed automation is applicable to any DPA and does not assume conformance with any predefined template.}

\section{Threats to Validity} \label{sec:threats}

Below, we discuss threats to the validity of our empirical results and what we did to mitigate these threats.

\sectopic{Internal Validity. } The main concern with respect to internal validity is bias. To mitigate this threat, we curated our evaluation dataset exclusively through third-parties (non-authors). The annotators were never exposed to the implementation details of our approach. Another potential threat is subjectivity in our interpretation of the GDPR text when extracting the compliance requirements related to DPA. To mitigate this threat, requirements extraction was done in close collaboration with three legal experts, who have expertise in European and international laws.  Our extracted requirements are further made publicly available and thus open to scrutiny. \textcolor{black}{Another threat to internal validity is our reliance on most-frequent word sense (MFS) disambiguation for identifying the semantically related words to enrich the texts of DPAs. 
Using more advanced disambiguation methods might improve the results of our approach. However, the wide use of MFS in the NLP literature and its reported performance provide reasonable confidence about the results of our approach. That said, further experimentation can help mitigate this threat.}  
\sectopic{External Validity. } 
We evaluated our approach on 30 real DPAs from different sectors. Our approach performed well when identifying violations related to the GDPR requirements. This provides us with reasonable confidence that our solution is generalizable. That said, experimenting with additional DPAs can help further mitigate support external validity.

\section{Extending our Methodology beyond GDPR}
\label{sec:reusability}

{\color{black}Our suggested procedure for assessing the compliance of DPAs can be applied on regulations beyond GDPR or document types other than DPAs. To illustrate this point, assume that a regulation is denoted as $\mathcal{G}$, and a legal document type is denoted as $\mathcal{T}$. In our context, GDPR is an instance of $\mathcal{G}$, and DPAs are instances of $\mathcal{T}$. To reuse the same methodology for other instances of $\mathcal{G}$ and $\mathcal{T}$, based on our experience, we anticipate the following steps and  effort needed for each step. 

\begin{enumerate}
    \item Requirements elicitation from $\mathcal{G}$ with respect to $\mathcal{T}$, consuming about 30\% of the effort. This step depends mainly on the availability of legal experts and the size and complexity of the provisions in $\mathcal{G}$. 
    \item Define semantic frames (SFs) over the compliance requirements derived from $\mathcal{G}$, consuming about 10\% of the effort. Prior to defining the SFs, one has to verify whether the same semantic roles (SRs) defined in our work can be re-used as-is in the new context. If not, SRs must be refined accordingly. Defining the SFs should preferably be performed  in collaboration with legal experts. Therefore, this step depends as well on their availability, though their degree of involvement would typically be much lower than in the previous step.  
    \item Curate an annotated dataset for $\mathcal{T}$, consuming about 10\% of the effort. The first thing to consider is whether (unlabeled) data for $\mathcal{T}$ are available. Curating a labeled dataset of examples from $\mathcal{T}$ is a prerequisite for developing any automation. The annotation task involves manually checking the compliance of the textual content of $\mathcal{T}$ against the requirements derived from $\mathcal{G}$. This step largely depends on data availability as well as the availability of annotators who have the right expertise. One should anticipate for providing training material about compliance with respect to $\mathcal{G}$.     
    \item Develop an automation strategy for checking the compliance of $\mathcal{T}$ against $\mathcal{G}$, consuming about 50\% of the effort.  The main things to adjust in our approach include the extraction rules of SRs, and the similarity thresholds applied for matching the predicates and arguments between the text of $\mathcal{T}$ and the  requirements of $\mathcal{G}$.
\end{enumerate}
}


\section{Conclusion} \label{sec:conclusion}

In this paper, we propose an automated approach for checking the compliance of data processing agreements (DPAs) against the general data protection regulation (GDPR). Such DPAs have a significant impact on the requirements of systems processing personal data. In close collaboration with legal experts, we extract and document DPA-related requirements as ``shall'' requirements. Such documentation provides a shared understanding between requirements engineers and legal experts. In our approach, we first manually create a representation of the GDPR requirements based on semantic frames (SFs). Our approach then automatically generates SF-based representations for the textual content of a DPA, and 
subsequently compares this representation against GDPR to check whether the DPA is compliant.
Our evaluation is based on 30 real DPAs, curated by trained third-party annotators with a strong background in law. \textcolor{black}{Over this evaluation dataset, our approach achieves an average accuracy of 84.6\%, a precision of 89.1\% and a recall of 82.4\%}. We also show how higher accuracy can be achieved by focused reviews on a small percentage of DPA statements. We compare our approach against a baseline that relies on an off-the-shelf NLP pipeline. Our approach outperforms this baseline with an average gain of $\approx$20 percentage points in accuracy. In the future, we plan to curate more DPAs and further investigate the applicability of machine learning methods for enabling automated compliance checking.

\ifCLASSOPTIONcompsoc
  \section*{Acknowledgments}
\else
  \section*{Acknowledgment}
\fi
This paper was supported by Linklaters, Luxembourg's National Research Fund (FNR) under grant BRIDGES/19/IS/13759068/ARTAGO, and NSERC of Canada under the Discovery and CRC programs. 
The authors would like to specially thank Katrien Baetens, Sylvie Forastier, Peter Goes and Catherine Freichel from Linklaters LLP for enriching this work through their expertise.

\ifCLASSOPTIONcaptionsoff
  \newpage
\fi

\bibliographystyle{IEEEtran}
\balance
\bibliography{paper}

\newpage
\onecolumn
\appendices
\section{DPA Compliance Requirements in GDPR}
\label{sec:appendix_1}

\begin{table*}[h]
\caption{\textcolor{black}{Optional DPA Compliance Requirements in GDPR.}}\label{tab:optional-requirements}
\vspace{-.5em}
 \footnotesize
  \centering
	  \begin{threeparttable}[t]
  \begin{tabularx}{0.98\textwidth} {@{} p{0.08\textwidth}  
  @{\hskip 0.5em}  X}
   \toprule
   ID (\textbf{Cat}\tnote{1}) & Requirement (\textit{Reference}\tnote{2}) \\ 
    \midrule
    R27 (\textbf{MD7}) & The  organizational and technical measures to ensure a level of security can include: (a) pseudonymization and encryption of personal data, (b) ensure confidentiality, integrity, availability and resilience of processing systems and services, (c) restore the availability and access to personal data in a timely manner in the event of a physical or technical incident, and (d) regularly testing, assessing and evaluating the effectiveness of technical and organizational measures for ensuring the security of the processing. (\textit{Art.~32(1)})\\  
    \midrule
    R28 (\textbf{MD8}) & The notification of personal data breach shall at least include (a) the nature of personal data breach; (b) the name and contact details of the data protection officer; (c) the consequences of the breach; (d) the measures taken or proposed to mitigate its effects. (\textit{Art.~33(3)}) \\
    \midrule
    R29 (\textbf{MD9}) & The DPIA shall at least include (a) a systematic description of the envisaged processing operations and the purposes of the processing, (b) an assessment of the necessity and proportionality of the processing operations in relation to the purposes, (c) an assessment of the risks to the rights and freedoms of data subjects, and (d) the measures envisaged to address the risks. (\textit{Art.~35(7)})\\
    \midrule
    R30 (\textbf{PO20}) & The processor shall not transfer personal data to a third country or international organization without  a prior specific or general authorization of the controller. (\textit{Art.~28(3)(a)}) \\
   \midrule
   R31 (\textbf{PO21})& The processor can demonstrate guarantees to Article 28 (1--4) through adherence to an approved codes of conduct or an approved certification mechanism. (\textit{Art.~28(5)}) \\
   \midrule
   R32 (\textbf{PO22})& The processor shall implement appropriate technical and organizational measures to ensure a level of security appropriate to the risk of varying likelihood and severity for the rights and freedoms of natural persons. (\textit{Art.~32(1)})\\
   \midrule
   R33 (\textbf{PO23})& The processor shall ensure that any natural person acting under its authority who has access to personal data only process them on instructions from the controller. (\textit{Art.~32(4)}) \\
   \midrule
   R34 (\textbf{PO24})& The processor shall notify the controller without undue delay after becoming aware of a personal data breach. (\textit{Art.~33(2)}) \\
   \midrule
   R35 (\textbf{PO25})& A processor shall be liable for the damage caused by processing only where it has not complied with GDPR obligations specifically directed to processors or where it has acted outside or contrary to lawful instructions of the controller. (\textit{Art.~82(2)}) \\ 
   \midrule
   R36 (\textbf{CO1}) & The controller shall  inform the supervisory authority no later than 72 hours after having become aware of a personal data breach,. (\textit{Art.~33(1)})\\ 
    \midrule
    R37 (\textbf{CO2})  & The controller shall document personal data breaches. (\textit{Art.~33(5)})\\ 
    \midrule
    R38 (\textbf{CO3}) & In case of high risks, the controller shall communicate the data breach to the data subject without undue delay. (\textit{Art.~34(1)})\\ 
    \midrule
    R39 (\textbf{CO4})  & The controller shall carry out a DPIA. (\textit{Art.~35(1)})\\ 
    \midrule
    R40 (\textbf{CO5}) & The controller shall seek advice of the DPO when carrying a DPIA. (\textit{Art.~35(2)})\\ 
    \midrule
    R41 (\textbf{CO6}) & The controller shall seek the views of data subjects or their representatives on the intended processing. (\textit{Art.~35(9)})\\ 
    \midrule
    R42 (\textbf{CO7}) & The controller shall carry out a review to assess if processing is performed in accordance with the data protection impact assessment at least when there is a change of the risk represented by processing operations. (\textit{Art.~35(11)})\\ 
    \midrule
    R43 (\textbf{CO8}) & A controller shall be liable for the damage caused by any processing infringing the GDPR. (\textit{Art.~82(2)})\\ 
    \midrule
    R44 (\textbf{CR2}) & The controller shall have the right to suspend the processing in certain cases. (\textit{Linklaters LLP})\\ 
    \midrule
    R45 (\textbf{CR3}) & The controller shall have the right to terminate the DPA in certain cases. (\textit{Linklaters LLP}) \\ 
   \bottomrule
 \end{tabularx}
 	\begin{tablenotes}
    \item[1] Cat is the category of the requirement, namely metadata (MD), processor's obligation (PO), controller's right (CR), and controller's obligation (CO).  
    \item[2] GDPR-related articles.
    \end{tablenotes}
    \end{threeparttable}
\end{table*}

\begin{table*}[h]
\caption{Glossary Terms for DPA Compliance Requirements in GDPR.}\label{tab:glossary}
\vspace{-.5em}
  \centering
   \begin{threeparttable}[t]
  \begin{tabularx}{0.98\textwidth} {@{} p{0.25\textwidth}  @{\hskip 0.5em}  X}
   \toprule
   Concept (\textit{Art.~}\tnote{1}) & Definition \\
   \toprule
   Personal Data (\textit{Art.~4(1)}) & Any information relating to an identified or identifiable natural person (data subject). \\
    \midrule
    Processing (\textit{Art.~4(2)}) & Any operation or set of operations which is performed on personal data or on sets of personal data, whether or not by automated means, such as collection, recording, organization, structuring, storage, adaptation or alteration, retrieval, consultation, use, disclosure by transmission, dissemination or otherwise making available, alignment or combination, restriction, erasure or destruction. \\
    \midrule
    Pseudonymization (\textit{Art.~4(5)}) & The processing of personal data in such a manner that the personal data can no longer be attributed to a specific data subject without the use of additional information, provided that such additional information is kept separately and is subject to technical and organizational measures to ensure that the personal data are not attributed to an identified or identifiable natural person. \\
    \midrule
    Data Controller (\textit{Art.~4(7)}) & A natural or legal person, public authority, agency or other body which, alone or jointly with others, determines the purposes and means of the processing of personal data; where the purposes and means of such processing are determined by Union or Member State law, the controller or the specific criteria for its nomination may be provided for by Union or Member State law. \\
    \midrule
    Data Processor (\textit{Art.~4(8)}) & A natural or legal person, public authority, agency or other body which processes personal data on behalf of the controller. \\
    \midrule
    Personal Data Breach (\textit{Art.~4(12)}) & A breach of security leading to the accidental or unlawful destruction, loss, alteration, unauthorized disclosure of, or access to, personal data transmitted, stored or otherwise processed. \\
    \midrule
    Supervisory Authority (\textit{Art.~4(22)}) & A supervisory authority which is concerned by the processing of personal data because the controller or processor is established on the territory of the Member State of that supervisory authority; data subjects residing in the Member State of that supervisory authority are substantially affected or likely to be substantially affected by the processing; or a complaint has been lodged with that supervisory authority. \\
    \midrule
    Objection  (\textit{Art.~4(24)}) & An objection to a draft decision as to whether there is an infringement of this Regulation, or whether envisaged action in relation to the controller or processor complies with this Regulation, which clearly demonstrates the significance of the risks posed by the draft decision as regards the fundamental rights and freedoms of data subjects and, where applicable, the free flow of personal data within the Union. \\
    \midrule
    International Organization \par (\textit{Art.~4(26)}) & An organization and its subordinate bodies governed by public international law, or any other body which is set up by, or on the basis of, an agreement between two or more countries. \\
    \midrule
    Sub-processor (\textit{Art.~28(4)}) & A natural person or organization engaged by a Processor for carrying out specific processing activities on behalf of the Controller. \\
    \midrule
    Security of Processing (\textit{Art.~32(1)}) & Taking into account the state of the art, the costs of implementation and the nature, scope, context and purposes of processing as well as the risk of varying likelihood and severity for the rights and freedoms of natural persons, the controller and the processor shall implement appropriate technical and organizational measures to ensure a level of security appropriate to the risk. \\
    \midrule
    Data Protection Impact Assessment \par(\textit{Art.~35(7)}) & An assessment of the necessity and proportionality of the processing operations in relation to the purposes, considering the risks to the rights, freedoms and legitimate interests of data subjects (abbreviated as DPIA). \\
    \midrule
    Codes of Conduct (\textit{Art.~40(1)}) & Are intended to contribute to the proper application of this Regulation, taking account of the specific features of the various processing sectors and the specific needs of micro, small and medium-sized enterprises. \\
    \midrule
    Certification (\textit{Art.~42(1)}) & The establishment of data protection certification mechanisms and of data protection seals and marks, for the purpose of demonstrating compliance with this Regulation of processing operations by controllers and processors. The specific needs of micro, small and medium-sized enterprises shall be taken into account. \\
    \midrule
    Personal Data Transfer (\textit{Art.~45(1)}) & A transfer of personal data to a third country or an international organization may take place where the Commission has decided that the third country, a territory or one or more specified sectors within that third country, or the international organization in question ensures an adequate level of protection. If agreed by the parties, for the transfer of personal data processed under this agreement from a third country or international organization, they can rely on different mechanisms, including the prior consent of the controller. \\
    \bottomrule
 \end{tabularx}
 \begin{tablenotes}
     \item[1] GDPR-related articles.
   \end{tablenotes}
 \end{threeparttable}
\end{table*}

you can choose not to have a title for an appendix
if you want by leaving the argument blank

\newpage
\onecolumn
\section{SF-based Representations of Mandatory
and Optional Requirements}\label{sec:appendix_2}
\begin{table*}[h]
\caption{\textcolor{black}{SF-based Representations of Mandatory GDPR Requirements (R7 -- R26 in Table~\ref{tab:mandatory-requirements})}}\label{tab:sf1}
 \footnotesize
  \centering
  \begin{threeparttable}[t]
  \begin{tabularx}{0.98\textwidth} {@{} p{0.04\textwidth} @{\hskip 0.5em}  X}
   \toprule
   \textbf{ID}\S & Representation $\langle \text{\textbf{p}}\rangle$,  $\mathcal{A}$ \\
   \toprule
   \textbf{PO1} & $\langle \text{\textbf{not engage}}\rangle$, $\mathcal{A}=\{a_i : 0\leq i \leq 3\}$ such that $a_0$~=~[the processor]$_\textit{actor}$, $a_1$~=~[a sub-processor]$_\textit{object}$, $a_2$~=~[without prior specific or general written approval of the controller]$_\textit{constraint}$, $a_3$~=~[prior specific or general written approval]$_\textit{time}$ \\
   \midrule
   \textbf{PO2} & $\langle \text{\textbf{inform}}\rangle$,  $\mathcal{A}=\{a_i : 0\leq i \leq 4\}$ such that $a_0$~=~[the processor]$_\textit{actor}$, $a_1$~=~[the controller]$_\textit{beneficiary}$, $a_2$~=~[any intended changes]$_\textit{object}$, $a_3$~=~[the addition or replacement of sub-processors]$_\textit{situation}$, $a_4$~=~[in case of written authorization]$_\textit{condition}$ \\
   \midrule
   \textbf{PO3}  & $\langle \text{\textbf{process}}\rangle$, $\mathcal{A}=\{a_i : 0\leq i \leq 2\}$ such that $a_0$~=~[the processor]$_\textit{actor}$, $a_1$~=~[personal data]$_\textit{object}$, $a_2$~=~[on documented instructions from the controller]$_\textit{constraint}$ \\
   \midrule
   \textbf{PO4} & $\langle \text{\textbf{inform}}\rangle$, $\mathcal{A}=\{a_i : 0\leq i \leq 5\}$ such that $a_0$~=~[the processor]$_\textit{actor}$, $a_1$~=~[that legal requirement]$_\textit{object}$, $a_2$~=~[the controller]$_\textit{beneficiary}$, $a_3$~=~[If the processor requires by Union or Member State law to process personal data without instructions and law does not prohibit informing the controller on grounds of public interest]$_\textit{condition}$, $a_4$~=~[before processing]$_\textit{time}$, , $a_5$~=~[Union or Member State law]$_\textit{reference}$\\
   \midrule
   \textbf{PO5} & $\langle \text{\textbf{ensure}}\rangle$, $\mathcal{A}=\{a_i : 0\leq i \leq 3\}$ such that $a_0$~=~[the processor]$_\textit{actor}$, $a_1$~=~[persons]$_\textit{object}$, $a_2$~=~[have committed themselves to confidentiality]$_\textit{situation}$, $a_3$~=~[are under an appropriate statutory obligation of confidentiality]$_\textit{constraint}$\\
   \midrule
   \textbf{PO6} & $\langle \text{\textbf{take}}\rangle$, $\mathcal{A}=\{a_i : 0\leq i \leq 3\}$ such that $a_0$~=~[the processor]$_\textit{actor}$, $a_1$~=~[all measures]$_\textit{object}$, $a_2$~=~[Article 32]$_\textit{reference}$, $a_3$~=~[to ensure the security of processing]$_\textit{reason}$ \\
   \midrule
   \textbf{PO7} & $\langle \text{\textbf{assist}}\rangle$, $\mathcal{A}=\{a_i : 0\leq i \leq 4\}$ such that $a_0$~=~[the processor]$_\textit{actor}$, $a_1$~=~[the controller]$_\textit{beneficiary}$, $a_2$~=~[in fulfilling its obligation]$_\textit{object}$, $a_3$~=~[to respond to requests for exercising the data subject’s rights]$_\textit{reason}$, $a_4$~=~[requests]$_\textit{situation}$ \\
   \midrule
   \textbf{PO8} & $\langle \text{\textbf{assist}}\rangle$, $\mathcal{A}=\{a_i : 0\leq i \leq 2\}$ such that $a_0$~=~[the processor]$_\textit{actor}$, $a_1$~=~[the controller]$_\textit{beneficiary}$, $a_2$~=~[in ensuring the security of processing]$_\textit{object}$ \\
   \midrule
   \textbf{PO9} & $\langle \text{\textbf{assist}}\rangle$, $\mathcal{A}=\{a_i : 0\leq i \leq 5\}$ such that $a_0$~=~[the processor]$_\textit{actor}$, $a_1$~=~[the controller]$_\textit{beneficiary}$, $a_2$~=~[in consulting the supervisory authorities]$_\textit{object}$, $a_3$~=~[prior to processing]$_\textit{time}$, $a_4$~=~[to mitigate the risk]$_\textit{reason}$, $a_5$~=~[where the processing would result in a high risk in the absence of measures taken by the controller]$_\textit{constraint}$ \\
   \midrule
   \textbf{PO10} & $\langle \text{\textbf{assist}}\rangle$, $\mathcal{A}=\{a_i : 0\leq i \leq 3\}$ such that $a_0$~=~[the processor]$_\textit{actor}$, $a_1$~=~[the controller]$_\textit{beneficiary}$, $a_2$~=~[in notifying to the supervisory authority]$_\textit{object}$, $a_3$~=~[a personal data breach]$_\textit{situation}$ \\
   \midrule
   \textbf{PO11} & $\langle \text{\textbf{assist}}\rangle$, $\mathcal{A}=\{a_i : 0\leq i \leq 3\}$ such that $a_0$~=~[the processor]$_\textit{actor}$, $a_1$~=~[the controller]$_\textit{beneficiary}$, $a_2$~=~[in communicating to the data subject]$_\textit{object}$, $a_3$~=~[a personal data breach]$_\textit{situation}$ \\
   \midrule
   \textbf{PO12} & $\langle \text{\textbf{assist}}\rangle$, $\mathcal{A}=\{a_i : 0\leq i \leq 2\}$ such that $a_0$~=~[the processor]$_\textit{actor}$, $a_1$~=~[the controller]$_\textit{beneficiary}$, $a_2$~=~[in ensuring compliance with the obligations pursuant to data protection impact assessment (DPIA)]$_\textit{object}$ \\
   \midrule
   \textbf{PO13} & $\langle \text{\textbf{return or delete}}\rangle$, $\mathcal{A}=\{a_i : 0\leq i \leq 4\}$ such that $a_0$~=~[the processor]$_\textit{actor}$, $a_1$~=~[all personal data]$_\textit{object}$, $a_2$~=~[the controller]$_\textit{beneficiary}$, $a_3$~=~[after $|$ end]$_\textit{time}$, $a_4$~=~[the end of the provision of services related to processing]$_\textit{situation}$ \\
   \midrule
   \textbf{PO14} & $\langle \text{\textbf{inform}}\rangle$, $\mathcal{A}=\{a_i : 0\leq i \leq 3\}$ such that $a_0$~=~[the processor]$_\textit{actor}$, $a_1$~=~[the controller]$_\textit{beneficiary}$, $a_2$~=~[if an instruction infringes the GDPR or other data protection provisions]$_\textit{condition}$,  $a_3$~=~[GDPR or other data protection provisions]$_\textit{reference}$\\
   \midrule
   \textbf{PO15} & $\langle \text{\textbf{make available}}\rangle$, $\mathcal{A}=\{a_i : 0\leq i \leq 2\}$ such that $a_0$~=~[the processor]$_\textit{actor}$, $a_1$~=~[the controller]$_\textit{beneficiary}$, $a_2$~=~[information necessary to demonstrate compliance with the obligations]$_\textit{object}$, $a_2$~=~[Article 28 in GDPR]$_\textit{reference}$ \\
   \midrule
   \textbf{PO16} & $\langle \text{\textbf{allow for and contribute to}}\rangle$, $\mathcal{A}=\{a_i : 0\leq i \leq 2\}$ such that $a_0$~=~[the processor]$_\textit{actor}$, $a_1$~=~[audits including inspections]$_\textit{object}$, $a_2$~=~[conducted by the controller or another auditor mandated by the controller]$_\textit{situation}$ \\
   \midrule
   \textbf{PO17} & $\langle \text{\textbf{impose}}\rangle$, $\mathcal{A}=\{a_i : 0\leq i \leq 4\}$ such that $a_0$~=~[the processor]$_\textit{actor}$, $a_1$~=~[the engaged sub-processors]$_\textit{beneficiary}$, $a_2$~=~[the same obligations]$_\textit{object}$, $a_3$~=~[by way of contract or other legal act under]$_\textit{constraint}$, $a_4$~=~[Union or Member State law]$_\textit{reference}$ \\
   \midrule
   \textbf{PO18} & $\langle \text{\textbf{remain fully liable}}\rangle$, $\mathcal{A}=\{a_i : 0\leq i \leq 2\}$ such that $a_0$~=~[the processor]$_\textit{actor}$, $a_1$~=~[the controller]$_\textit{beneficiary}$, $a_2$~=~[for the performance of sub-processor’s obligations]$_\textit{object}$ \\
   \midrule
   \textbf{PO19} & $\langle \text{\textbf{take into account}}\rangle$, $\mathcal{A}=\{a_i : 0\leq i \leq 2\}$ such that $a_0$~=~[the processor]$_\textit{actor}$, $a_1$~=~[the risk of]$_\textit{object}$, $a_2$~=~[accidental or unlawful destruction, loss, alternation, unauthorized disclosure of or access to the personal data transmitted, stored or processed]$_\textit{situation}$ \\
   \midrule
   \textbf{CR1} & $\langle \text{\textbf{have the right}}\rangle$, $\mathcal{A}=\{a_i : 0\leq i \leq 5\}$ such that $a_0$~=~[the controller]$_\textit{actor}$, $a_1$~=~[to object to changes]$_\textit{object}$, $a_2$~=~[in case of written authorization]$_\textit{condition}$, $a_3$~=~[by the processor]$_\textit{constraint}$, $a_4$~=~[after having been informed of such intended changes]$_\textit{time}$, $a_5$~=~[addition or replacement of sub-processors]$_\textit{situation}$ \\
   \bottomrule
 \end{tabularx}
 \end{threeparttable}
\end{table*}

\begin{table*}[h]
\caption{\textcolor{black}{SF-based Representations of Optional GDPR requirements (R27 -- R45 in Table~\ref{tab:optional-requirements})}}\label{tab:sf2}
 \footnotesize
  \centering
  \begin{threeparttable}[t]
  \begin{tabularx}{0.98\textwidth} {@{} p{0.04\textwidth} @{\hskip 0.5em}  X}
   \toprule
   \textbf{ID}\S & Representation $\langle \text{\textbf{p}}\rangle$,  $\mathcal{A}$ \\
   \toprule
   \textbf{PO20} & $\langle \text{\textbf{not transfer}}\rangle$, $\mathcal{A}=\{a_i : 0\leq i \leq 4\}$ such that $a_0$~=~[the processor]$_\textit{actor}$, $a_1$~=~[a third country or international organization]$_\textit{beneficiary}$, $a_2$~=~[personal data]$_\textit{object}$, $a_3$~=~[without a prior specific or general authorization of the controller]$_\textit{constraint}$, $a_4$~=~[prior specific or general authorization]$_\textit{time}$ \\
   \midrule
    \textbf{PO21} & $\langle \text{\textbf{demonstrate}}\rangle$, $\mathcal{A}=\{a_i : 0\leq i \leq 3\}$ such that $a_0$~=~[the processor]$_\textit{actor}$, $a_1$~=~[guarantees]$_\textit{object}$, $a_2$~=~[adherence to an approved code of conduct or an approved certification mechanism]$_\textit{situation}$, $a_3$~=~[Article 28 (1-4)]$_\textit{reference}$ \\
    \midrule
    \textbf{PO22} & $\langle \text{\textbf{implement}}\rangle$, $\mathcal{A}=\{a_i : 0\leq i \leq 3\}$ such that $a_0$~=~[the processor]$_\textit{actor}$, $a_1$~=~[appropriate technical and organizational measures]$_\textit{object}$, $a_2$~=~[to ensure a level of security appropriate to the risk]$_\textit{reason}$, $a_3$~=~[varying likelihood and severity for the rights and freedoms of natural persons]$_\textit{situation}$ \\
    \midrule
    \textbf{PO23} & $\langle \text{\textbf{ensure}}\rangle$, $\mathcal{A}=\{a_i : 0\leq i \leq 3\}$ such that $a_0$~=~[the processor]$_\textit{actor}$, $a_1$~=~[any person]$_\textit{beneficiary}$, $a_2$~=~[under its authority who has access to personal data acts only on instructions from the controller]$_\textit{constraint}$, $a_3$~=~[access to personal data]$_\textit{situation}$ \\
    \midrule
    \textbf{PO24} & $\langle \text{\textbf{notify}}\rangle$, $\mathcal{A}=\{a_i : 0\leq i \leq 3\}$ such that $a_0$~=~[the processor]$_\textit{actor}$, $a_1$~=~[the controller]$_\textit{beneficiary}$, $a_2$~=~[without undue delay]$_\textit{constraint}$, $a_3$~=~[a data breach]$_\textit{situation}$ \\
    \midrule
    \textbf{PO25} & $\langle \text{\textbf{be liable}}\rangle$, $\mathcal{A}=\{a_i : 0\leq i \leq 3\}$ such that $a_0$~=~[a processor]$_\textit{actor}$, $a_1$~=~[for the damage caused]$_\textit{object}$, $a_2$~=~[where acting outside or contrary to lawful instructions of the controller or not complying with obligations of the GDPR specifically directed to processors]$_\textit{condition}$, $a_3$~=~[by processing]$_\textit{constraint}$ \\
    \midrule
    \textbf{CO1} & $\langle \text{\textbf{inform}}\rangle$, $\mathcal{A}=\{a_i : 0\leq i \leq 3\}$ such that $a_0$~=~[the controller]$_\textit{actor}$, $a_1$~=~[the supervisory authority]$_\textit{beneficiary}$, $a_2$~=~[no later than 72 hours after having become aware]$_\textit{time}$, $a_3$~=~[a personal data breach]$_\textit{situation}$ \\
    \midrule
    \textbf{CO2} & $\langle \text{\textbf{document}}\rangle$, $\mathcal{A}=\{a_i : 0\leq i \leq 1\}$ such that $a_0$~=~[the controller]$_\textit{actor}$, $a_1$~=~[personal data breaches]$_\textit{object}$ \\
    \midrule
    \textbf{CO3} & $\langle \text{\textbf{inform}}\rangle$, $\mathcal{A}=\{a_i : 0\leq i \leq 4\}$ such that $a_0$~=~[the controller]$_\textit{actor}$, $a_1$~=~[the data subject]$_\textit{beneficiary}$, $a_2$~=~[in case of high risks]$_\textit{condition}$, $a_3$~=~[without undue delay]$_\textit{constraint}$, $a_4$~=~[a data breach]$_\textit{situation}$ \\
    \midrule
    \textbf{CO4} & $\langle \text{\textbf{carry out}}\rangle$, $\mathcal{A}=\{a_i : 0\leq i \leq 1\}$ such that $a_0$~=~[the controller]$_\textit{actor}$, $a_1$~=~[DPIA]$_\textit{object}$ \\
    \midrule
    \textbf{CO5} & $\langle \text{\textbf{seek}}\rangle$, $\mathcal{A}=\{a_i : 0\leq i \leq 2\}$ such that $a_0$~=~[the controller]$_\textit{actor}$, $a_1$~=~[advice of the DPO]$_\textit{object}$, $a_2$~=~[when carrying out DPIA]$_\textit{condition}$ \\
    \midrule
    \textbf{CO6} & $\langle \text{\textbf{seek}}\rangle$, $\mathcal{A}=\{a_i : 0\leq i \leq 1\}$ such that $a_0$~=~[the controller]$_\textit{actor}$, $a_1$~=~[the views of data subjects or their representatives on the intended processing]$_\textit{object}$ \\
    \midrule
    \textbf{CO7} & $\langle \text{\textbf{carry out}}\rangle$, $\mathcal{A}=\{a_i : 0\leq i \leq 5\}$ such that $a_0$~=~[the controller]$_\textit{actor}$, $a_1$~=~[a review]$_\textit{object}$, $a_2$~=~[at least when there exists]$_\textit{condition}$, $a_3$~=~[represented by processing operations]$_\textit{constraint}$, $a_4$~=~[to assess if processing is performed in accordance with the DPIA]$_\textit{reason}$, $a_5$~=~[a change of the risk]$_\textit{situation}$ \\
    \midrule
    \textbf{CO8} & $\langle \text{\textbf{be liable}}\rangle$, $\mathcal{A}=\{a_i : 0\leq i \leq 3\}$ such that $a_0$~=~[a controller]$_\textit{actor}$, $a_1$~=~[for the damage]$_\textit{object}$, $a_2$~=~[caused by any processing infringing the GDPR]$_\textit{constraint}$, $a_3$~=~[GDPR]$_\textit{reference}$ \\
    \midrule
    \textbf{CR2} & $\langle \text{\textbf{have the right}}\rangle$, $\mathcal{A}=\{a_i : 0\leq i \leq 2\}$ such that $a_0$~=~[the controller]$_\textit{actor}$, $a_1$~=~[to suspend the processing]$_\textit{object}$, $a_2$~=~[in certain cases]$_\textit{condition}$ \\
    \midrule
    \textbf{CR3} & $\langle \text{\textbf{have the right}}\rangle$, $\mathcal{A}=\{a_i : 0\leq i \leq 3\}$ such that $a_0$~=~[the controller]$_\textit{actor}$, $a_1$~=~[to terminate]$_\textit{object}$, $a_2$~=~[in certain cases]$_\textit{condition}$, $a_3$~=~[the DPA]$_\textit{reference}$ \\
   \bottomrule
 \end{tabularx}

 \end{threeparttable}
\end{table*}

\end{document}